\documentclass[onecolumn,noshowpacs,superscriptaddress,nobibnotes,nofootinbib,12pt,showkeys,preprintnumbers]{revtex4-1}
\UseRawInputEncoding
\usepackage{float}
\usepackage{amsmath,amssymb}
\usepackage{bm}
\usepackage{slashed}
\usepackage{epsfig}
\usepackage{graphicx}
\usepackage{hyperref}
\usepackage{xcolor}
\hypersetup{
    colorlinks=true,
    linkcolor=blue,
    filecolor=magenta,      
    urlcolor=blue,
    citecolor=blue
}

\usepackage[utf8]{inputenc}
\usepackage{placeins}
\usepackage{braket}
\usepackage{subfigure}
\graphicspath{ {image/} }

\newcommand*\diff{\mathop{}\!\mathrm{d}}

\newcommand{\Si}{\textrm{Si}}
\newcommand{\Ci}{\textrm{Ci}}
\newcommand{\GeV}{{{\,}\textrm{GeV}}}

\colorlet{darkgreen}{green!50!black}
\colorlet{brightyellow}{yellow!75!red}
\colorlet{orange}{red!50!yellow}
\colorlet{darkgray}{gray!50!black}
\colorlet{darkred}{red!50!black}

\usepackage{soul}

\definecolor{lcolor}{rgb}{0.5,0,0}
\definecolor{citcolor}{rgb}{0,0.3,0.0}

\begin{document}
\title{Anisotropic flow and the valence quark skeleton of hadrons}
\author{Meijian Li}
\email{meijian.li@usc.es}
\author{Wenyang Qian}
\email{qian.wenyang@usc.es}
\author{Bin Wu}
\email{b.wu@cern.ch}
\affiliation{Instituto Galego de F\'isica de Altas Enerx\'ias IGFAE, Universidade de Santiago de Compostela,
E-15782 Galicia-Spain}
\author{Hong Zhang}
\email{hong.zhang@sdu.edu.cn}
\affiliation{Institute of Frontier and Interdisciplinary Science, Key Laboratory of Particle
Physics and Particle Irradiation (MOE), Shandong University, Qingdao,
Shandong 266237, China
}

\begin{abstract}
We study transverse momentum anisotropies, in particular, the elliptic flow $v_2$ due to the interference effect sourced by valence quarks in high-energy hadron-hadron collisions. Our main formula is derived as the high-energy (eikonal) limit of the impact-parameter dependent cross section in quantum field theory, which agrees with that in terms of the impact parameter in the classical picture. As a quantitative assessment of the interference effect, we calculate $v_2$ in the azimuthal distribution of gluons at a comprehensive coverage of the impact parameter and the transverse momentum in high-energy pion-pion collisions. In a broad range of the impact parameter, a sizable amount of $v_2$, comparable with that produced due to saturated dense gluons or final-state interactions, is found to develop. In our calculations, the valence sector of the pion wave function is obtained numerically from the Basis Light-Front Quantization, a non-perturbative light-front Hamiltonian approach. And our formalism is generic and can be applied to other small collision systems like proton-proton collisions.
\end{abstract}

\maketitle

\section{Introduction}

In heavy-ion collisions, azimuthal anisotropies have long been viewed as one of the major signatures for the formation of strongly-coupled quark-gluon plasma (QGP) fluid droplets~\cite{Heinz:2013th}. The unexpected observations of collectivity in proton-proton (pp) and proton-nucleus (pA) collisions~\cite{CMS:2010ifv, CMS:2012qk, ALICE:2012eyl, ATLAS:2012cix, CMS:2013jlh, ALICE:2014dwt, CMS:2014und, CMS:2015yux, ATLAS:2015hzw, CMS:2015fgy, LHCb:2015coe, CMS:2016est, CMS:2016fnw, ATLAS:2017hap, CMS:2017xnj, ATLAS:2017rtr, CMS:2017kcs, PHENIX:2018lia, ATLAS:2018ngv, CMS:2019fur, ATLAS:2019wzn, CMS:2020qul} have, however, put such a QGP signature under intense scrutiny and inspired extensive exploration of other potential sources for collectivity in the past decade~(see refs.~\cite{Nagle:2018nvi, Altinoluk:2020wpf} for recent reviews). Pinning down the true origin of collectivity in small collision systems is one of the major goals to be pursued by the heavy-ion community in the decades to come~\cite{Citron:2018lsq}.

Extending the hydrodynamic paradigm in heavy-ion collisions to small collision systems has yielded some phenomenological success in explaining experimental data as pioneered in refs.~\cite{dEnterria:2010xip, Bozek:2010pb, Bozek:2011if}. Its applicability has thus far been justified to some extent by studies in microscopic models which concluded the dominance of hydrodynamic modes in systems comparable to, or even considerably smaller than, the proton size as summarized in \cite{Romatschke:2017ejr}. On the other hand, it is yet to be understood how hydrodynamization could be established in such small systems from QCD first principles. Moreover, sizable momentum anisotropies can already be produced via one final-state interaction as shown in kinetic theory~\cite{Borghini:2010hy, Romatschke:2018wgi, Kurkela:2018ygx, Kurkela:2018qeb, Kurkela:2021ctp} or via color reconnection in PYTHIA~\cite{OrtizVelasquez:2013ofg}, which all indicates that hydrodynamics, if present, may not be solely responsible for collectivity in small collision systems~\cite{Greif:2017bnr, Kurkela:2019kip, Kurkela:2020wwb, Lin:2021mdn, Ambrus:2022qya}.

Without considering any final-state interactions, transverse momentum anisotropies could also be produced in pp and pA collisions. Various initial-state correlations responsible for such momentum anisotropies have been intensively investigated in parton saturation physics, i.e., the Color-Glass Condensate (CGC)~\cite{Armesto:2006bv, Dumitru:2010iy, Kovner:2010xk, Levin:2011fb, Kovner:2011pe, Dusling:2012iga, Kovchegov:2012nd, Dusling:2013oia, Kovchegov:2013ewa, Schenke:2014zha, Dumitru:2014yza, Altinoluk:2015uaa, Lappi:2015vta, Schenke:2016lrs, Kovner:2016jfp, Iancu:2017fzn, Dusling:2017dqg, Dusling:2017aot, Kovchegov:2018jun, Mace:2018vwq, Altinoluk:2018ogz, Mace:2018yvl, Davy:2018hsl, Agostini:2019hkj, Agostini:2019avp, Agostini:2021xca, Agostini:2022ctk, Agostini:2022oge}. Many of these studies (see, e.g., \cite{Kovner:2016jfp, Iancu:2017fzn, Dusling:2017dqg, Dusling:2017aot, Kovchegov:2018jun, Mace:2018vwq, Altinoluk:2018ogz, Mace:2018yvl, Davy:2018hsl, Agostini:2019hkj, Agostini:2019avp, Agostini:2021xca, Agostini:2022ctk, Agostini:2022oge}) were focused on the dilute-dense limit in which only one of the two colliding particles is modelled as a saturated dense gluon state  while the other (the proton) is treated as dilute. As an alternative mechanism  without explicitly resorting to high initial parton saturation, interference between gluon waves emitted by various classical sources in hadrons has been shown to be capable of generating significant azimuthal asymmetries (in higher-order cumulants) as well~\cite{Blok:2017pui, Blok:2018xes}\footnote{The approach to the interference effect in these works, otherwise, shares some commonalities with the glasma graph approach in CGC~\cite{Altinoluk:2015uaa}.}.

Most of the above studies suffer from uncertainties due to some ad hoc modeling of color sources (large-$x$ partons)\footnote{See ref.~\cite{Dumitru:2020gla} and references therein for recent progress on this issue.}. We note that the needed information on color sources or multi-parton distributions could also be acquired from hadron wave functions evaluated in the following approaches: The light-front Hamiltonian formalism describes the hadron internal structure and dynamics through the light-front wavefunctions~(LFWFs), which are fully relativistic and nonperturbative~\cite{Brodsky:1997de}. The basis light-front quantization~(BLFQ) approach emerges as a computational framework to solve bound-state problems by employing basis representations~\cite{Vary:2009gt}. The hadron LFWFs are obtained by diagonalizing the phenomenological effective Hamiltonian operator, which is based on light-front holography~\cite{deTeramond:2008ht,Brodsky:2014yha,Brodsky:2020ajy}. The application of BLFQ ranges from heavy mesons~\cite{Li:2015zda,Li:2017mlw,Tang:2018myz,Li:2021cwv}, heavy-light meson~\cite{Tang:2019gvn}, light meson~\cite{Qian:2020utg,Jia:2018ary,Zhu:2023lst,Li:2022izo,Lan:2022blr} and baryons~\cite{Liu:2022fvl,Hu:2022ctr,Xu:2022dbw,Xu:2022abw}. The obtained LFWFs have provided us with opportunities to study various observables and physical processes.

In this paper we derive a formalism for studying high-energy hadron-hadron collisions from the impact-parameter dependent cross section in quantum field theory as defined in \cite{Wu:2021ril} and show that sizable transverse momentum anisotropies (mainly $v_2$) in soft gluon production can be produced due to the interference effect induced by the emitters of valence (anti)quarks in the dilute-dilute limit. Below, we explain the physical picture behind our calculations and summarize our main results.

The  physical picture that motivates this work is as follows. Since we are mostly interested in low-$p_T$ final-state hadrons which are not part of jets initiated by high-$p_T$ partons, the relevant final-state partons (mostly gluons) typically have a transverse momentum $p_T\sim1$~GeV and, equivalently, a de Broglie wavelength $\lambda$ comparable with the transverse size of the colliding hadrons ($\sim1$~fm). Accordingly, their production is expected to be sensitive to the incoming color-singlet state of multipartons that are typically separated by a transverse distance $\sim\lambda$. In this case we expect that the most relevant multiparton state is given by the hadron valence quark skeleton as broadly assumed in parton saturation physics~\cite{Kovchegov:2012mbw}. Motivated by the above expectation, we investigate below soft gluon production in the collisions of hadron valence quark skeletons as described by the LFWFs.

In sec.~\ref{sec:b_cross_section}, we present the main formula eq.~(\ref{eq:dsigmadbdo_pipi}) and the Feynman rules eq.~(\ref{eq:FeynRulesFinal}) for calculating, order by order in perturbation theory, the gluon production in impact-parameter dependent hadron-hadron collisions. This formula is derived by taking the high collision energy (eikonal) limit of the impact-parameter dependent cross section defined in \cite{Wu:2021ril}, as transcribed below in  eq.~(\ref{eq:dsigmadbdosym}). In this limit the dipole cross section for meson-meson collisions agrees with that broadly used in saturation/small-$x$ physics~\cite{Mueller:1993rr,Mueller:1994jq, Mueller:1994gb, Kovchegov:2005ur}. This provides another verification of eq.~(\ref{eq:dsigmadbdosym}) as a generic definition for the impact-parameter dependent cross section in quantum field theory, complementing to the hard (Drell-Yan) process studied in~\cite{Wu:2021ril}.

In sec.~\ref{sec:dipole_dipole} we derive the meson-meson (mainly the dipole-dipole) cross section at leading order: first for inelastic forward scattering and then for one gluon production. With the latter result presented in eq.~(\ref{eq:sigppg}), we are able to study the momentum anisotropies of soft gluon production. In our gauge choice in which one meson does not radiate (to the midrapidity region), the cross section encodes the interference pattern of the two rays of gluon radiated respectively from the valence $q\bar{q}$ in the other meson. It is analogous to, though more sophisticated than, the classical double-slit experiment in optics, as discussed in detail in sec.~\ref{sec:dipole_dipole_v2}. With such an analog in mind, this work shares some similarity with refs.~\cite{Blok:2017pui, Blok:2018xes} as well as ref.~\cite{Altinoluk:2015uaa}, which dealt with a more intricate interference pattern among multiple gluons from multiple emitters.

Our main results of $v_2$ for pion-pion collisions follows in sec.~\ref{sec:pion_pion} (see fig.~\ref{fig:v2kT}). With the only two parameters in the meson LFWFs fixed by light meson masses as briefly reviewed in sec.~\ref{sec:wavefunction}, the results of $v_2$ at different impact parameters are shown and discussed in sec.~\ref{sec:pionv2}. The main observations include: 1) Without any free parameters (except the impact parameter $b$) the magnitude of $v_2$ in the gluon production is typically comparable to that observed in pp collisions~\cite{ATLAS:2015hzw, CMS:2016fnw, ATLAS:2017hap, ATLAS:2017rtr, CMS:2017kcs, ATLAS:2018ngv, ATLAS:2019wzn, CMS:2020qul} as well as theoretical results~\cite{dEnterria:2010xip, Bozek:2010pb, Habich:2015rtj, Weller:2017tsr, Zhao:2020pty, Dumitru:2010iy, Dusling:2012iga, Dusling:2013oia, Schenke:2014zha, Schenke:2016lrs, Iancu:2017fzn, Altinoluk:2020wpf}. 2) The interference pattern is found to manifest as a distinct, double-peak structure in $v_2$ for $b\gtrsim 0.1$~fm, which encodes the information on the hadron size.

\section{The impact-parameter dependent cross section}\label{sec:b_cross_section}

Following refs.~\cite{Levin:2011fb, Iancu:2017fzn}, we investigate how transverse momentum anisotropies in the gluon production are correlated with the impact parameter of high-energy hadron-hadron collisions. We instead focus on the dilute-dilute limit, meaning both hadrons are taken as color-singlet multiparton states instead of saturated gluon states. In this section we derive the impact-parameter dependent cross section for producing partons of transverse momentum $p_T$ in the limit $p_T\ll\sqrt{s}$, the center-of-mass (CM) energy of hadron-hadron collisions. 

\subsection{The impact-parameter dependent cross section in the eikonal approximation}

In quantum field theory the impact-parameter dependent cross section for the collision of two high-energy particles can be unambiguously defined as follows if the fuzziness in beam particles’ transverse positions is much smaller than the impact parameter $b=|\mathbf{b}|$~\cite{Wu:2021ril}
\begin{align}\label{eq:dsigmadbdosym}
        \frac{d\sigma}{d^2{\mathbf b} dO}     =&\int\prod\limits_{i=A,B}\frac{d^2\mathbf{q}_i}{(2\pi)^2}e^{-i\mathbf{q}_i\cdot\mathbf{x}_i}\int\prod\limits_f\left[d\Gamma_{p_f}\right]\delta(O-O(\{p_f\}))(2\pi)^2\delta^{(2)}(\mathbf{\tilde{P}}_A + \mathbf{\tilde{P}}_B - \sum \mathbf{p}_f)\notag\\
        &\times\frac{1}{2s} M(P_A, P_B \to \{p_f\}) M^*(\tilde{P}_A, \tilde{P}_B \to \{p_f\})(2\pi)^4\delta^{(4)}(P_A + P_B - \sum p_f),
\end{align}
where $O(\{p_f\})$ defines an observable $O$ as a function of the final-state momenta $\{p_f\}$, $M(P_A, P_B \to \{p_f\})$ is the amplitude for the process: $P_A, P_B \to \{p_f\}$, the phase-space measure for a particle of momentum $p_f$ and mass $m_f$ is defined as
\begin{align}\label{eq:Gammapf}
    \int d\Gamma_{p_f}\equiv \int \frac{d^4 p_f}{(2\pi)^4}(2\pi)\delta(p_f^2-m_f^2)\theta(p_f^0),
\end{align}
in the CM frame the incoming momenta are given by
\begin{align}\label{eq:psincomingsym}
    P_i^\mu = \frac{\sqrt{s}}{2}n_i^\mu+\frac{q_i^{\mu}}{2},~~\tilde{P}_i^\mu = \frac{\sqrt{s}}{2}n_i^\mu - \frac{q_i^{\mu}}{2}
\end{align}
with 
\begin{align}
    n_A^\mu = (1, 0, 0, 1),
     \qquad n_B^{\mu}=(1, 0, 0, -1),\qquad\;q_i^\mu = (0,\mathbf{q}_i,0)
\end{align}
and ${\mathbf{x}}_i$ can be identified with the transverse location of hadron $i$ in the classical picture with the impact parameter ${\mathbf{b}}\equiv{\mathbf{x}}_A-{\mathbf{x}}_B$.
Here, all the terms of $O(q^2)$ or $O(m_{i}^2)$ with $i=A,B$ in the incoming momenta are neglected and two-dimensional transverse vectors are denoted by boldface letters.

For the production of high-$p_T$ particles the underlying process is typically initiated by a binary collision of two partons respectively collinear to the two beam directions. In QCD this picture is manifest in the limit when the transverse coherent length of the collision $\sim1/p_T$ is much shorter than the average separation between the partons in the hadrons $\sim 1/\Lambda_{QCD}$. In this case the cross section for the process takes a factorized form when expanded to the leading order in $1/p_T$, i.e., at leading twist. And the only information of the hadron substructure encoded in such a hard process is, in general, the transverse phase-space distributions of single partons, referred to as thickness beam functions in \cite{Wu:2021ril}.

Complementing to hard processes, we focus on the production of relatively low-$p_T$ particles near midrapidity in the CM frame. At $p_T\sim 1$~GeV, the de Broglie wavelength of the produced particles $\lambda=2\pi/p$ becomes comparable with the transverse size of the colliding hadrons. Accordingly, one may expect that the production of low-$p_T$ particles is not sensitive to one but multipartons that are typically separated by a distance $\sim\lambda$ in the transverse plane inside the colliding hadrons. To substantiate such an expectation, below we make a detailed calculation of the azimuthal angle distribution of soft gluons produced in hadron-hadron collisions. In order to assess the sole importance of such an effect, we shall neglect other sources of azimuthal momentum anisotropies as recently reviewed in~\cite{Nagle:2018nvi,Altinoluk:2020wpf}.

We resort to the light-front wave functions (LFWFs) to decompose hadron states into color-singlet multiparton states~\cite{Brodsky:1997de} and focus on the valence sector of the hadron wave functions\footnote{As broadly taken in parton saturation physics~\cite{Kovchegov:2012mbw}, this approximation is motivated by the observation that the valence quarks dominate at large $x$. The validity of such an approximation could be verified quantitatively by introducing one more collinear gluon in the Fock state (see, e.g., \cite{Lan:2022blr} for mesons and \cite{Dumitru:2020gla, Xu:2022abw} for the proton).
}. In this case our calculations boil down to the evaluation of the partonic cross section for soft gluon production in the collision of two color singlet states of valence (anti)quarks. Unlike the expansion of Feynman diagrams in $p_T^{-1}$ for hard processes, we expand all the graphs in $s^{-1}$ and keep only leading-order terms in $s^{-1}$. This corresponds to the so-called eikonal limit. The eikonal approximation has been broadly used in the Glauber model for heavy-ion collisions~\cite{Miller:2007ri} and parton saturation physics~\cite{Kovchegov:2012mbw}. Under this approximation we simplify the expression for the impact-parameter dependent cross section in eq.~(\ref{eq:dsigmadbdosym}) and include below all ensuing Feynman rules for self-containment.

Our calculations are to be carried out in a mixed representation where longitudinal components of momenta and transverse coordinates are used to describe single particle states. Each valence (anti)quark carries a finite momentum fraction of its parent hadron and moves predominantly along the hadron's lightlike direction, denoted by $n$. In this case, the longitudinal components of their momentum $p$ can be conveniently chosen to be $\bar{n}\cdot\;p$ and $n\cdot\;p$ with $\bar{n}^\mu\equiv(1,-\vec{n})$, which are respectively called the plus (``+") and minus (``-") components of $p$ along $n$. That is, by definition a valence (anti)quark always carries a large plus momentum $\propto\sqrt{s}$ along $n$. By expanding in its plus momentum, Feynman rules associated to the valence (anti)quark in the mixed representation reduce to
\begin{align}
\label{eq:FeynRules}
&\begin{array}{lll}
\text{External quark lines:}&\includegraphics[width=0.07\textwidth]{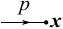}=u^s_n e^{i\mathbf{p}\cdot\mathbf{x}}&
\includegraphics[width=0.07\textwidth]{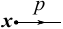}=\bar{u}^s_n e^{-i\mathbf{p}\cdot\mathbf{x}}
\\
\text{External antiquark lines:}&\includegraphics[width=0.07\textwidth]{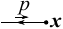}=\bar{v}^s_n e^{i\mathbf{p}\cdot\mathbf{x}}&
\includegraphics[width=0.07\textwidth]{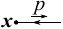}=v^s_n e^{-i\mathbf{p}\cdot\mathbf{x}}
\\
\text{External gluon lines:}&\includegraphics[width=0.07\textwidth]{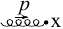}=\epsilon_\lambda(p) e^{i\mathbf{p}\cdot\mathbf{x}}&
\includegraphics[width=0.07\textwidth]{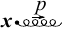}=\epsilon^*_\lambda(p) e^{-i\mathbf{p}\cdot\mathbf{x}}
\\
\text{Quark-gluon vertex:}& \includegraphics[width=0.1\textwidth]{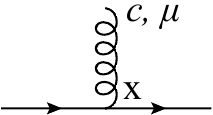} = -ig t^c\dfrac{\slashed{\bar{n}}}{2}n^\mu\int\,d^2\mathbf{x}
\end{array}\\
&\begin{array}{lll}
\text{Gluon propagator:}&\includegraphics[width=0.1\textwidth]{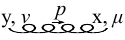}=\int\frac{d^2\mathbf{k}}{(2\pi)^2}\frac{-i g^{\mu\nu}e^{i\mathbf{k}\cdot (\mathbf{x}-\mathbf{y})}}{n\cdot\,p\bar{n}\cdot p - |\mathbf{k}|^2+i\epsilon}\\
\text{Quark propagator:}&\includegraphics[width=0.1\textwidth]{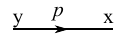}= \frac{i\frac{\slashed{n}}{2}\delta^{(2)}(\mathbf{x}-\mathbf{y})}{n\cdot\,p+i\bar{n}\cdot\,p~\epsilon}
~~~\includegraphics[width=0.1\textwidth]{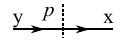}
= \frac{\slashed{n}}{2}(2\pi)\delta(n\cdot\,p)\delta^{(2)}(\mathbf{x}-\mathbf{y}).&
\end{array}\notag
\end{align}
In addition both the ``+" and ``-" momenta along $n$ are conserved at each vertex. Since the momenta of valence (anti)quarks do not enter the observable $O$ (near midrapidity), they are integrated out, giving rise to the delta function for the above cut fermion line. Here, the spinors with subscript $n$ are defined for momentum $p^\mu = \frac{\bar{n}\cdot p}{2}n^\mu$.

Given a generic cut graph with the above Feynman rules, one can further make the following simplifications.  
Let us single out a cut fermion line corresponding to one valence (anti)quark:
\begin{align}\label{eq:Sq}
    S_q\equiv
    \begin{array}{c}
    \includegraphics[width=0.2\textwidth]{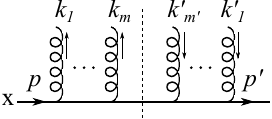}
    \end{array}.
\end{align}
First, according to longitudinal momentum conservation and the Feynman rule for the cut fermion line, $S_q$ contains the following delta functions:
\begin{align}\label{eq:Sqdeltapm}
  2(2\pi)\delta(\bar{n}\cdot p - \bar{n}\cdot k-\bar{n}\cdot p')(2\pi)\delta^-(\sum\limits_{i=1}^{m}{n}\cdot k_i)(2\pi)\delta^-(\sum\limits_{i=1}^{m'}{n}\cdot k'_i),
\end{align}
where $k$ is the total momentum of gluons in both the amplitude and the conjugate amplitude and the function $\delta^-$ is defined as
\begin{align}\label{eq:dminus}
\delta^-(\sum\limits_{i=1}^m{n}\cdot k_i)=\left\{
\begin{array}{cc}
1/(2\pi)&\text{for $m=0$}  \\
\delta(\sum\limits_{i=1}^m{n}\cdot k_i)& \text{for $m\geq\,1$}
\end{array}
\right.
.
\end{align}
In the limit $\bar{n}\cdot\;k_i,\bar{n}\cdot\;k'_i
\ll\bar{n}\cdot\;p$, the difference between $\bar{n}\cdot p$ and $\bar{n}\cdot p'$ in the spinors and Dirac matrices can be ignored. Accordingly, they can be simplified as follows
\begin{align}\label{eq:Sqspinors}
   \bar{u}_n^{s'}\frac{\slashed{\bar{n}}}{2}\frac{\slashed{n}}{2}\cdots\frac{\slashed{\bar{n}}}{2}\frac{\slashed{n}}{2}\frac{\slashed{\bar{n}}}{2}u_n^s=\bar{u}_n^{s'}\frac{\slashed{\bar{n}}}{2}u_n^s=(\bar{n}\cdot\,p)\delta^{s's}.
\end{align}
Besides, the transverse coordinates of the valence (anti)quark remain the same due to the delta function in the quark propagator.

Using the above results to simplify the rules in eq.~(\ref{eq:FeynRules}) yields:

{1. Feynman rules}
\begin{align}
\label{eq:FeynRulesFinal}
&\begin{array}{lll}
\text{External valence $q/\bar{q}$:}&\includegraphics[width=0.07\textwidth]{image/qin.pdf}=1
&
\includegraphics[width=0.07\textwidth]{image/qbin.pdf}=1
\\
\text{External gluon lines:}&\includegraphics[width=0.07\textwidth]{image/gin.pdf}=\epsilon_\lambda(p) e^{i\mathbf{p}\cdot\mathbf{x}}&
\includegraphics[width=0.07\textwidth]{image/gout.pdf}=\epsilon^*_\lambda(p) e^{-i\mathbf{p}\cdot\mathbf{x}}
\\
\text{Quark-gluon vertex:}& \includegraphics[width=0.1\textwidth]{image/qqg.pdf} = -ig t^cn^\mu\vspace{2mm}
\\
\text{Gluon propagator:}&\includegraphics[width=0.1\textwidth]{image/gg.pdf}=\int\frac{d^2\mathbf{k}}{(2\pi)^2}\frac{-i g^{\mu\nu}e^{i\mathbf{k}\cdot (\mathbf{x}-\mathbf{y})}}{n\cdot\,p\bar{n}\cdot p - |\mathbf{k}|^2+i\epsilon}
\end{array}\\
&\begin{array}{lll}
\text{valence quark propagator:}&\includegraphics[width=0.1\textwidth]{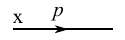}= \frac{i}{n\cdot\,p+i\bar{n}\cdot\,p~\epsilon}&
~~~~\includegraphics[width=0.1\textwidth]{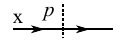}
= 1.
\end{array}\notag
\end{align}

2. For an initial-state (anti)quark, insert a phase factor $e^{i\mathbf{p}\cdot\mathbf{x}}$ or $e^{-i\mathbf{p}\cdot\mathbf{x}}$ respectively in the amplitude or the conjugate amplitude. Here, $\mathbf{p}$ and $\mathbf{x}$ are respectively the transverse momentum and the transverse coordinates of the (anti)quark.

3. Impose the conservation of ``+" momentum and spin across each (anti)quark line in the cut diagrams:
\begin{align}\label{eq:FeynRulesCoefs}
   \Delta^{s's}_+(p,p',k)\equiv\;&2\bar{n}\cdot\,p \delta^{s's}(2\pi)\delta(\bar{n}\cdot p - \bar{n}\cdot k-\bar{n}\cdot p')
\end{align}
with $k$ the total momentum of gluons.

4. Impose ``-" momentum conservation for gluons hooked on the (anti)quark line respectively in the amplitude and the conjugate amplitude as long as the number of gluons is not zero. For a cut diagram with $m$ and $m'$ gluons in the amplitude and the conjugate amplitude respectively, one has
\begin{align}\label{eq:deltam2}
(2\pi)\delta^-(\sum\limits_{i=1}^{m}{n}\cdot k_i)(2\pi)\delta^-(\sum\limits_{i=1}^{m'}{n}\cdot k'_i)
\end{align}
with $\delta^-$ defined in eq.~(\ref{eq:dminus}).

5. Integrate over transverse coordinates of (anti)quark lines and undetermined longitudinal momenta.
\hspace{2mm}

\subsection{The impact-parameter dependent cross section for pion-pion collisions}

 Using the above Feynman rules, one can write down explicitly the expression for the impact-parameter dependent cross section in the collisions between baryons (color-singlet triples) and/or mesons (color-singlet dipoles). Since the pion LFWFs are among the most studied ones~\cite{Qian:2020utg}, we focus on high-energy pion-pion collisions in the rest of this paper.

As detailed in sec.~\ref{sec:wavefunction}, the incoming pions are taken as valence $q\bar{q}$ states. Here, we need the LFWFs for mesons respectively moving along the two beam directions. Given one of the light-like vectors $n^\mu$, the LFWF is expressed as 
\begin{align}\label{eq:oniaStates}
\begin{split}
|\Psi_{h}(P)\rangle
=&\frac{1}{\sqrt{N_c}}\sum_{c=1}^{N_c}\sum\limits_{s_q, s_{\bar{q}},\xi}\int\limits_{\mathbf{p}}\psi_{s_q\;s_{\bar{q}}/h}(\mathbf{p},\xi) |p_q, c, s_q; p_{\bar{q}},c,s_{\bar{q}}\rangle
  \;,
\end{split}
\end{align}
where $s_q$ and $s_{\bar{q}}$ are the quark and antiquark spins respectively, $c$ is the color index, the longitudinal momentum fraction $\xi\equiv \bar n\cdot p_q/\bar n\cdot P$ and in terms of the quark (antiquark) momentum $p_q$ ($p_{\bar{q}}$) the pion transverse momentum $\mathbf{P}$ and the relative momentum $\mathbf{p}$ are respectively defined as
\begin{align}\label{eq:pTqqb}
\mathbf{P}=\mathbf{p}_q+\mathbf{p}_{\bar{q}},\qquad\mathbf{p}=(1-\xi) \mathbf{p}_q-\xi\mathbf{p}_{\bar{q}}.
\end{align}
Here, for brevity we have dropped some quantum numbers that are to be restored in sec.~\ref{sec:wavefunction}  and used the following shorthand notation
\begin{align}
    \sum\limits_{s, \bar{s},\xi}\equiv\sum\limits_{ s_q, s_{\bar{q}}}\frac{1}{4\pi}\int_0^1\frac{d\xi}{{\xi(1-\xi)}},\qquad \int\limits_{\mathbf{p}}\equiv\int \frac{d^2\mathbf{p}}{(2\pi)^2}.
\end{align}

With pions taken as valence $q\bar{q}$ states, the evaluation of pion-pion cross sections is boggled down to that for the scattering of color-singlet dipoles. The impact-parameter dependent cross section in eq.~(\ref{eq:dsigmadbdosym}) can be decomposed into two parts respectively associated with the two pions. Each part can be extracted from
\begin{align}\label{eq:pion2dipole}
S_{\pi}\equiv&\int\limits_{\mathbf{q}}\frac{e^{-i\mathbf{q}\cdot\mathbf{x_{\pi}}}}{\bar{n}\cdot\,P}\left(
    \begin{array}{c}
        \includegraphics[width=0.25\textwidth]{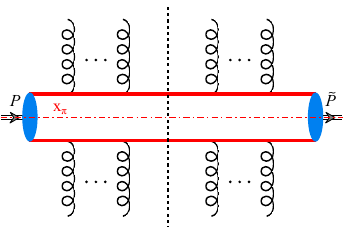}
    \end{array}\right),
\end{align}
where the cut diagram stands for the $S$-matrix element squared, $\mathbf{x}_{\pi}$ is the pion transverse coordinates and $\mathbf{P}=-\mathbf{\tilde{P}}=\mathbf{q}/2$. Note that the number of gluons in the amplitude can be different from that in the conjugate amplitude since each gluon field can be either contracted with another one in the same $S_{\pi}$ or in the other. Plugging the pion LFWF into the above equation gives
\begin{align}
S_{\pi}=&\frac{1}{N_c}\int\limits_{\mathbf{x}_{q},\mathbf{x}_{\bar{q}}}\sum\limits_{s,\bar{s},\xi}\sum\limits_{s',\bar{s}',\xi'}\int\limits_{\mathbf{q},\mathbf{p}, \mathbf{\tilde{p}}}{e^{-i\mathbf{q}\cdot\mathbf{x_{\pi}}+i(\mathbf{p}_{q}-\mathbf{\tilde{p}}_{q})\cdot\mathbf{x_{q}}+i(\mathbf{p}_{\bar{q}}-\mathbf{\tilde{p}}_{\bar{q}})\cdot\mathbf{x}_{\bar{q}}}}\psi_{s\bar{s}/h}(\mathbf{p},\xi)\psi^*_{s'\,\bar{s}'/h}(\mathbf{\tilde{p}},\xi')\notag\\
&\times\frac{1}{\bar{n}\cdot\,P}\Delta_+^{s's}(p_q,\tilde{p}_q,k)\Delta_+^{\bar{s}'\bar{s}}(p_{\bar{q}},\tilde{p}_{\bar{q}},\bar{k})
\left(
    \begin{array}{c}
        \includegraphics[width=0.25\textwidth]{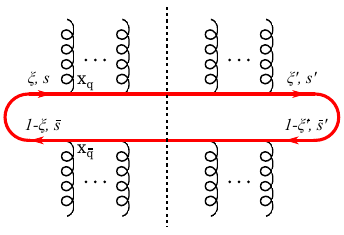}
    \end{array}\right),\notag
\end{align}
where $k$ and $\bar{k}$ are the total momenta of gluons respectively on the $q$ and $\bar{q}$ lines, the quark ($p_q$) and antiquark ($p_{\bar{q}}$) momenta in the amplitude are related to the pion momentum $P$ and $\mathbf{p}$ according to eq.~(\ref{eq:pTqqb}) while their momenta in the conjugate amplitude, denoted by an overhead tilde, are related to the pion momentum $\tilde{P}$, defined in eq.~(\ref{eq:psincomingsym}), and $\mathbf{\tilde{p}}$ in the same way. Here, the shorthand notation
\begin{align}
    \int\limits_{\mathbf{x}}\equiv \int\,d^2\mathbf{x}
\end{align}
is employed and the cut graph for the dipole in the last line is given by the Feynman rules in eq.~(\ref{eq:FeynRulesFinal}).

Let us further simplify the expression of $S_{\pi}$. First, one has
\begin{align}\label{eq:deltasdipole}
    &\frac{1}{\bar{n}\cdot\,P}\Delta_+^{s's}(p_q,\tilde{p}_q,\{k\})\Delta_+^{\bar{s}'\bar{s}}(p_{\bar{q}},\tilde{p}_{\bar{q}},\{\bar{k}\})= 4\pi\xi(1-\xi)\delta(\xi-\xi')\delta^{s\,s'}\delta^{\bar{s}\,\bar{s}'}4\pi\delta(\bar{n}\cdot\,k+\bar{n}\cdot\bar{k})
\end{align}
according to the definition of $\Delta_+$ in eq.~(\ref{eq:FeynRulesCoefs}). Second, in terms of the pion CM and relative transverse coordinates
\begin{align}\label{eq:Xr}
    \mathbf{X}\equiv\xi \mathbf{x}_q+(1-\xi )\mathbf{x}_{\bar{q}},\qquad \mathbf{r}=\mathbf{x}_q-\mathbf{x}_{\bar{q}},
\end{align}
one has
\begin{align}\label{eq:phasesdipole}
    e^{i(\mathbf{p}_q-\mathbf{\tilde{p}}_q)\cdot\mathbf{x}_q+i(\mathbf{p}_{\bar{q}}-\mathbf{\tilde{p}}_{\bar{q}})\cdot\mathbf{x}_{\bar{q}}}=e^{i\mathbf{q}\cdot\,\mathbf{X} + i(\mathbf{p}-\mathbf{\tilde{p}})\cdot\mathbf{r}}
\end{align}
with $\xi$ and $\xi'$ identified. Accordingly, one has
\begin{align}\label{eq:Spi}
S_\pi=\int\limits_{\mathbf{r}}\sum\limits_{s_q,s_{\bar{q},\xi}}|\tilde\psi_{s_qs_{\bar{q}}/\sigma}(\mathbf{r},\xi)|^2\frac{1}{N_c}4\pi\delta(\bar{n}\cdot\,k+\bar{n}\cdot\bar{k})
\left(
    \begin{array}{c}
        \includegraphics[width=0.25\textwidth]{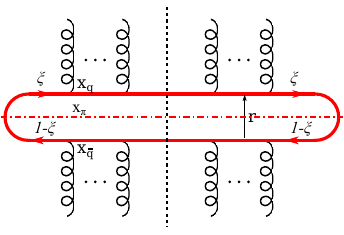}
    \end{array}\right),
\end{align}
where the wave function in the mixed representation is given by Fourier transform:
\begin{align}
\tilde\psi(\mathbf{r},\xi)\equiv\int \frac{d^2\mathbf{p}}{(2\pi)^2} e^{i \mathbf{p}\cdot \mathbf{r}} \frac{\psi(\mathbf{p},\xi)}{\sqrt{\xi(1-\xi)}}.
\end{align}

Finally, by removing the overall delta functions in  the two $S_\pi$'s for $n=n_A$ and $n_B$ and contracting all the gluon fields one has
\begin{align}
\label{eq:dsigmadbdo_pipi}
    \frac{d\sigma}{d^2{\mathbf b} dO}=&\prod\limits_{i=A,B}\int\,d^2{\mathbf{r}_i}\sum\limits_{s^i_q,s^i_{\bar{q}}}\frac{1}{4\pi}\int_0^1{d\xi_i}|\psi_{s^i_q\,s^i_{\bar{q}}/\sigma}(\mathbf{r}_i,\xi_i)|^2\frac{d\hat{\sigma}}{d^2{\mathbf b} dO},
\end{align}
where the dipole cross section at fixed impact parameter $\mathbf{b}$ is defined as
\begin{align}\label{eq:dsigmadbdo_dipole}
 \frac{d\hat{\sigma}}{d^2{\mathbf b} dO}\equiv\int\prod\limits_f\left[d\Gamma_{p_f}\right]\delta(O-O(\{p_f\}))\frac{1}{N_c^2}\left(
    \begin{array}{c}
        \includegraphics[width=0.3\textwidth]{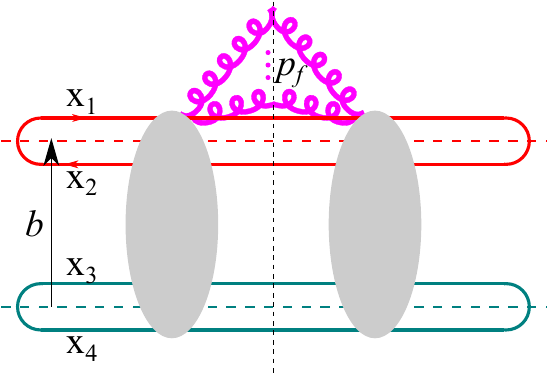}
    \end{array}\right),
\end{align}
in which the transverse coordinates of the valence quarks and antiquarks of the two pions are given by 
\begin{align}\label{eq:xi_to_br}
\begin{split}
&\mathbf{x}_1=\mathbf{x}_{A,q}=\mathbf{x}_{A}+(1-\xi_A) \mathbf{r}_A,\qquad
\mathbf{x}_2=\mathbf{x}_{A,\bar q}=\mathbf{x}_{A}-\xi_A\mathbf{r}_A,\\
&\mathbf{x}_3=\mathbf{x}_{B,q}=\mathbf{x}_{B}+(1-\xi_B) \mathbf{r}_B,\qquad
\mathbf{x}_4=\mathbf{x}_{B,\bar q}=\mathbf{x}_{B}-\xi_B\mathbf{r}_B.
\end{split}
\end{align}
That is, the CM transverse coordinates of the two pions are identified with $\mathbf x_{A}$ and $\mathbf x_{B}$ respectively. As confirmed in the next section, the impact-parameter dependent cross section in eq.~(\ref{eq:dsigmadbdo_dipole}) is the same as that broadly used in parton saturation/small-$x$ physics~\cite{Mueller:1993rr,Mueller:1994jq, Mueller:1994gb, Kovchegov:2005ur}. The above derivation hence confirms the generality of eq.~(\ref{eq:dsigmadbdosym}) as the definition of the impact-parameter dependent cross section in quantum field theory.

\section{Fixed-order impact-parameter dependent cross sections}\label{sec:dipole_dipole}
In this section we carry out detailed calculations of the impact-parameter dependent cross section and the azimuthal distribution for one gluon production in dipole-dipole collisions at leading order (LO).

\subsection{The impact-parameter dependent cross section at LO}
\label{sec:full_dipole_sig}
For the total impact-parameter dependent cross section, there are 16 cut diagrams corresponding to the 16 ways to connect the two dipoles with two gluon lines, each in the amplitude and the conjugate amplitude\footnote{
Here, we only consider inelastic scattering in which the  final-state $q\bar{q}$ pairs are not in color singlet. 
}. By using the Feynman rules in eq.~(\ref{eq:FeynRulesFinal}), one has
\begin{align}
    \begin{array}{c}
        \includegraphics[width=0.12\textwidth]{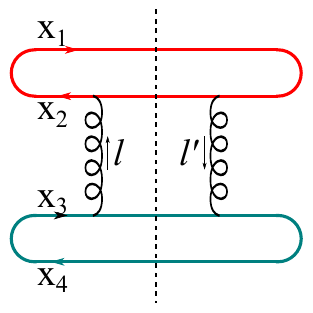}
    \end{array}=&\frac{g^4}{2}C_F N_c\int\frac{d^4l}{(2\pi)^4}\frac{in_A\cdot\,n_B}{l^2}e^{i\mathbf{l}\cdot(\mathbf{x}_2-\mathbf{x}_3)}\,2\pi\delta(n_A\cdot\,l)2\pi\delta(n_B\cdot\,l)\notag\\
    &\times\int\frac{d^4{l}'}{(2\pi)^4}\frac{-in_A\cdot\,n_B}{{l}'^2}e^{-i\mathbf{l}'\cdot(\mathbf{x}_2-\mathbf{x}_3)}\,2\pi\delta(n_A\cdot\,l')2\pi\delta(n_B\cdot\,l')\notag\\
    =&\frac{g^4}{2}C_F N_c\int\frac{d^2\mathbf{l}}{(2\pi)^2}\frac{e^{i\mathbf{l}\cdot(\mathbf{x}_2-\mathbf{x}_3)}}{|\mathbf{l}|^2}\int\frac{d^4\mathbf{l}'}{(2\pi)^2}\frac{e^{-i\mathbf{l}'\cdot(\mathbf{x}_2-\mathbf{x}_3)}}{|\mathbf{l}'|^2}.
\end{align}
And, including all possible hookings of the two gluons leads to
\begin{align}\label{eq:sigqqb}
  \frac{d\hat\sigma}{d^2\mathbf{b}}=&\frac{g^4 C_F}{2N_c}\bigg|\int\frac{d^2\mathbf{l}}{(2\pi)^2}\frac{1}{|\mathbf{l}|^2}\bigg(e^{i\mathbf{l}\cdot\mathbf{x}_1}-e^{i\mathbf{l}\cdot\mathbf{x}_2}\bigg)\bigg(e^{-i\mathbf{l}\cdot\mathbf{x}_3}-e^{-i\mathbf{l}\cdot\mathbf{x}_4}\bigg)\bigg|^2.
\end{align}
By using the integral
\begin{align}\label{eq:F1}
    F_1(x,y)\equiv\mu^{2-d}\int\frac{d^d \mathbf{l}}{(2\pi)^d} \frac{e^{i\mathbf{l}\cdot \mathbf{x}}}{|\mathbf{l}|^2}=\frac{1}{4\pi}\frac{\Gamma(-\epsilon)}{(\pi r^2 \mu^2)^{-\epsilon}} =-\frac{1}{4\pi}\bigg[\frac{1}{\epsilon} +\gamma_E+\ln (\pi r^2\mu^2)\bigg]\;
\end{align}
with $d=2-2\epsilon$, one finally has
\begin{align}\label{eq:sigLO}
  \frac{d\sigma}{d^2\mathbf{b}}
  =&\frac{2\alpha_s^2 C_F}{N_c}\prod\limits_{i=A,B}\int\,d^2{\mathbf{r}_i}\sum\limits_{s^i_q,s^i_{\bar{q}}}\frac{1}{4\pi}\int_0^1\frac{d\xi_i}{\xi_i(1-\xi_i)}\notag\\
  &\times|\psi_{s^i_q\,s^i_{\bar{q}}/\sigma}(\mathbf{r}_i,\xi_i)|^2\ln^2\left(\frac{|\mathbf{x}_1-\mathbf{x}_4||\mathbf{x}_2-\mathbf{x}_3|}{|\mathbf{x}_1-\mathbf{x}_3||\mathbf{x}_2-\mathbf{x}_4|}\right).
\end{align}
We, hence, confirm the known result in the literature (see, e.g., eq.~(3.139) in ref.~\cite{Kovchegov:2012mbw}).

\subsection{Soft gluon production in dipole-dipole scattering}

The impact-parameter dependent dipole cross section for radiating one gluon is given by
\begin{align}\label{eq:dsdb}
\frac{d\hat\sigma}{d^2\mathbf{b}d\eta d^2\mathbf{k}}=&\frac{1}{2 (2\pi)^3}\frac{1}{N_c^2}\left(
\begin{array}{c}\includegraphics[width=0.2\textwidth]{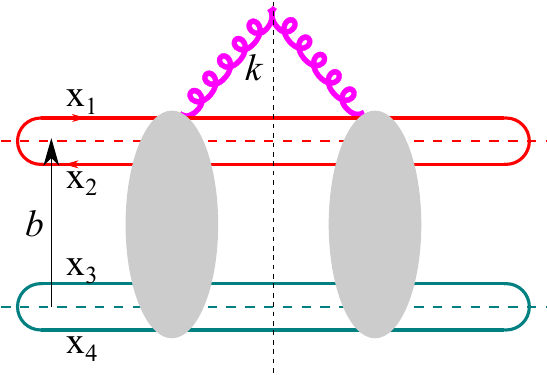}\end{array}\right)
\end{align}
with $\eta$ the (pseudo)rapidity of the gluon.

The above dipole cross section is gauge invariant and we choose to use $n_B\cdot\,A=0$ light-cone gauge in which the gluon polarization vector $\epsilon^\mu_\lambda(k)$ is given by
\begin{align}\label{eq:polarization}
    n_B\cdot\epsilon_\lambda(k)=0,\qquad\,k\cdot \epsilon_\lambda(k)=0\Rightarrow\,n_A\cdot\epsilon_\lambda(k) = \frac{2}{n_B\cdot\,k}\mathbf{\epsilon}_\lambda\cdot\mathbf{k}.
\end{align}
For observables near midrapidity, one only needs to consider diagrams with the soft gluon attached to the dipole of pion $A$ in this gauge~\cite{Wu:2017rry}. One can also discard diagrams in which the gluon is radiated without scattering. In such diagrams one has $n_A\cdot\;k=0$, that is, the gluon is moving along $n_A$. 

Using the above facts, the evaluation of the LO diagrams boils down to the replacement
\begin{align}\label{eq:onegluon}
 \begin{array}{c}\includegraphics[height=0.1\textheight]{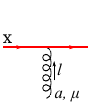}
 \end{array}
 \to
  \begin{array}{c}\includegraphics[height=0.1\textheight]{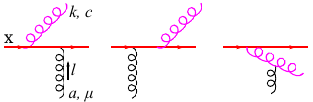}
  \end{array}
\end{align}
with the fermion line standing for the quark or antiquark in pion $A$. Contracting both sides of the above equation with $n_B$ and $\epsilon^*_\lambda(k)$ yields
\begin{align}\label{eq:liptov}
    -ig\,n_A\cdot\,n_B\,t^ae^{i\mathbf{l}\cdot\mathbf{x}}\to -i2g^2\,n_A\cdot\,n_B\,[t^c, t^a]\bigg(\frac{\mathbf{k}}{|\mathbf{k}|^2}-\frac{\mathbf{k}-\mathbf{l}}{|\mathbf{k}-\mathbf{l}|^2}\bigg)\cdot\mathbf{\epsilon}^*_{\lambda}(k)e^{i(\mathbf{l}-\mathbf{k})\cdot\mathbf{x}}
\end{align}
for the quark in the amplitude, which is also true for the antiquark up to an overall minus sign. 
And the corresponding replacement for the color factor after squaring the amplitude is
\begin{align}
     \text{Tr}(t^at^b)\text{Tr}(t^at^b) =\frac{C_F N_c}{2}\to \text{Tr}(t^at^b)\text{Tr}([t^c, t^a][t^a, t^c])=\frac{C_FN_c^2}{2}.
\end{align}

Making the above replacements in the total dipole cross section in eq.~(\ref{eq:sigqqb}), we finally obtain
\begin{align}\label{eq:sigppg}
  \frac{d\hat\sigma}{d^2\mathbf{b} d\eta\,d^2 \mathbf{k} }=&\frac{2\alpha_s^2 C_F}{N_c} 
  \int \frac{d^2 \mathbf{l}}{(2\pi)^2}\frac{1}{|\mathbf{l}|^2}\frac{d^2 \mathbf{l}'}{(2\pi)^2}\frac{1}{|\mathbf{l}'|^2}\left(e^{-i\mathbf{l}\cdot \mathbf{x}_3}-e^{-i\mathbf{l}\cdot \mathbf{x}_4}\right)\left(e^{i\mathbf{l}'\cdot \mathbf{x}_3} -e^{i\mathbf{l}'\cdot \mathbf{x}_4}\right)\notag\\
  &\times 4 \alpha_s N_c\left(\frac{\mathbf{k}}{|\mathbf{k}|^2}-\frac{\mathbf{k}-\mathbf{l}}{|\mathbf{k}-\mathbf{l}|^2}\right)\cdot\left(\frac{\mathbf{k}}{|\mathbf{k}|^2}-\frac{\mathbf{k}-\mathbf{l}'}{|\mathbf{k}-\mathbf{l}'|^2}\right)\notag\\
  &\times \left[ e^{i(\mathbf{l}-\mathbf{k})\cdot \mathbf{x}_1}-e^{i(\mathbf{l}-\mathbf{k})\cdot \mathbf{x}_2}\right]\left[e^{-i(\mathbf{l}'-\mathbf{k})\cdot \mathbf{x}_1} -e^{-i(\mathbf{l}'-\mathbf{k})\cdot \mathbf{x}_2} \right],
\end{align}
which is independent of $\eta$ and, therefore, longitudinally boost invariant due to the fact that the soft gluon spectrum is independent of the energies of the valence (anti)quarks.

The above dipole cross section respects the following symmetries. First, it is obviously invariant under $q\leftrightarrow\bar{q}$ in either dipole, i.e., $\mathbf{x}_1\leftrightarrow\mathbf{x}_2\text{ or }\mathbf{x}_3\leftrightarrow\mathbf{x}_4$. Second, it is invariant under $\mathbf{x}_1\leftrightarrow\mathbf{x}_3$ and, simultaneously, $\mathbf{x}_2\leftrightarrow\mathbf{x}_4$, as required by gauge invariance. This is not evident on the amplitude level since the relevant diagrams admit the interpretation as soft gluon radiation by the emitters of valence $q$ and $\bar{q}$ of pion $A$ in $n_B\cdot\,A=0$ gauge. In $n_A\cdot\,A=0$ gauge,  the dipole cross section is, instead, given by that in $n_B\cdot\,A=0$ gauge with $\mathbf{x}_1\leftrightarrow\mathbf{x}_3$ and $\mathbf{x}_2\leftrightarrow\mathbf{x}_4$. The equivalence of these two results can be straightforwardly verified by using the following relation
\begin{align}
    &\frac{2}{|\mathbf{l}|^2|\mathbf{l}'|^2}\left({|\mathbf{k}|^2}-\frac{\mathbf{k}-\mathbf{l}}{|\mathbf{k}-\mathbf{l}|^2}\right)\cdot\left(\frac{\mathbf{k}}{|\mathbf{k}|^2}-\frac{\mathbf{k}-\mathbf{l}'}{|\mathbf{k}-\mathbf{l}'|^2}\right)\notag\\
    &=\frac{1}{|\mathbf{l}|^2 |\mathbf{k}|^2||\mathbf{l}'-\mathbf{k}|^2}+\frac{1}{|\mathbf{l}'|^2 |\mathbf{k}|^2||\mathbf{l}-\mathbf{k}|^2}-\frac{|\mathbf{l}-\mathbf{l}'|^2}{|\mathbf{l}|^2 |\mathbf{l}'|^2||\mathbf{l}-\mathbf{k}|^2|\mathbf{l}'-\mathbf{k}|^2}.
\end{align}

\subsection{Evaluation of the dipole cross section for soft gluon production}
\label{eq:eval_sig_dipole}

Let us write the dipole cross section eq.~(\ref{eq:sigppg}) in the following form
\begin{align}\label{eq:sigppgJ}
  \frac{d\hat\sigma}{d^2\mathbf{b} d\eta d^2 \mathbf{k} }=&{8\alpha_s^3 C_F} |\mathbf{J}(\{\mathbf{x}_i\})|^2
\end{align}
with
\begin{align}\label{eq:J}
    \mathbf{J}(\{\mathbf{x}_i\})\equiv&
    \int \frac{d^2 \mathbf{l}}{(2\pi)^2}\frac{1}{|\mathbf{l}|^2}\left(e^{-i\mathbf{l}\cdot \mathbf{x}_3}-e^{-i\mathbf{l}\cdot \mathbf{x}_4}\right)\left(\frac{\mathbf{k}}{|\mathbf{k}|^2}-\frac{\mathbf{k}-\mathbf{l}}{|\mathbf{k}-\mathbf{l}|^2}\right)\left[ e^{i(\mathbf{l}-\mathbf{k})\cdot \mathbf{x}_1}-e^{i(\mathbf{l}-\mathbf{k})\cdot \mathbf{x}_2}\right].
\end{align}
In terms of the integrals
\begin{align}\label{eq:F1_euler}
\begin{split}
    F_1(x,y)\equiv\int\frac{d^2 \mathbf{l}}{(2\pi)^2} \frac{e^{i\mathbf{l}\cdot \mathbf{x}}}{|\mathbf{l}|^2},\qquad
    F_2(x,y)\equiv\int\frac{d^2 \mathbf{l}}{(2\pi)^2} \frac{e^{i\mathbf{l}\cdot \mathbf{x}}}{|\mathbf{l}|^2|\mathbf{k}-\mathbf{l}|^2}
    \;,
\end{split}
\end{align}
$\mathbf{J}$ can be expressed as
\begin{align}\label{eq:J_Fs}
    \mathbf{J}(\{\mathbf{x}_i\})
    &=\frac{\mathbf{k}}{|\mathbf{k}|^2}\bigg\{\left[F_1(\mathbf{x}_{13})-F_1(\mathbf{x}_{14})\right]e^{-i\mathbf{k}\cdot \mathbf{x}_1}-\left[F_1(\mathbf{x}_{23})-F_1(\mathbf{x}_{24})\right]e^{-i\mathbf{k}\cdot \mathbf{x}_2}\bigg\}\notag\\
  &-e^{-i\mathbf{k}\cdot \mathbf{x}_1}(\mathbf{k}+i \mathbf{\nabla})\left[F_2(\mathbf{x}_{13})-F_2(\mathbf{x}_{14})\right]
    +e^{-i\mathbf{k}\cdot \mathbf{x}_2}(\mathbf{k}+i \mathbf{\nabla})\left[F_2(\mathbf{x}_{23})-F_2(\mathbf{x}_{24})\right]\;,
\end{align}
where $\mathbf{x}_{ij} \equiv\mathbf{x}_{i}-\mathbf{x}_{j}$, and $ \nabla=(\partial_x, \partial_y)$ acts on the two-dimensional coordinate space.

Both $F_1$ and $F_2$ are singular while $\mathbf{J}$ is finite. One can further single out the divergent piece of $F_2$ by using the following relation
\begin{align}
    F_2 = \frac{1+e^{i\mathbf{k}\cdot\mathbf{x}}}{|\mathbf{k}|^2}F_1 - \frac{2}{|\mathbf{k}|^2}F_3,
\end{align}
where the finite integral $F_3$ is defined as
\begin{align}
    F_3(\mathbf{x})\equiv\int\frac{d^2 \mathbf{l}}{(2\pi)^2} e^{i\mathbf{l}\cdot \mathbf{x}}\frac{ \mathbf{l}\cdot(\mathbf{l}-\mathbf{k})}{|\mathbf{l}|^2|\mathbf{k}-\mathbf{l}|^2}.
\end{align}

$F_1$, regularized by dimensional regularization, is given in eq.~(\ref{eq:F1}) and here we only need to evaluate $F_3$. Let us choose a frame with the basis vectors given by
\begin{align}
    \mathbf{e}_1 = \frac{\mathbf{k}-\frac{\mathbf{k}\cdot \mathbf{x}}{r}\frac{\mathbf{x}}{r}}{\left|\mathbf{k}-\frac{\mathbf{k}\cdot \mathbf{x}}{r}\frac{\mathbf{x}}{r}\right|},\qquad \mathbf{e}_2 = \frac{\mathbf{x}}{r}.
\end{align}
In this frame, one has
\begin{align}
    \mathbf{x}=(x_1, x_2)=(0, r),\qquad
    \mathbf{k}=(k_1,k_2)=\bigg(\left|\mathbf{k}-\frac{\mathbf{k}\cdot \mathbf{x}}{r}\frac{\mathbf{x}}{r}\right|, \frac{\mathbf{k}\cdot \mathbf{x}}{r}\bigg),
\end{align}
and the integration over $l_1$ can be easily carried out by using the residue theorem with the poles given by
\begin{align}
    l_1 = \pm i l_2, \qquad k_1 \pm i (l_2-k_2).
\end{align}
Then, making a Fourier transform of $l_2$ yields
\begin{align}
F_3(r,\mathbf{k})=&\frac{1}{8 \pi }e^{-\frac{1}{2} r ({k_1}-i {k_2})} \bigg\{-e^{{k_1} r} \bigg[\text{Ci}\left(\frac{1}{2} (i{k_1} - {k_2}) r\right)+\text{Ci}\left(\frac{1}{2} (-i{k_1}- {k_2}) r\right)\notag\\
&-i\text{Si}\left(\frac{1}{2} (-i{k_1}+ {k_2}) r\right)+i \text{Si}\left(\frac{1}{2} (i {k_1}+{k_2}) r\right)\bigg]-\text{Ci}\left(\frac{1}{2} ({k_2}-i {k_1}) r\right)\notag\\
&-\text{Ci}\left(\frac{1}{2} (i {k_1}+{k_2}) r\right)-i\text{Si}\left(\frac{1}{2} (-i{k_1}+ {k_2}) r\right)+i\text{Si}\left(\frac{1}{2} (i{k_1}+{k_2}) r\right)\bigg\},
\end{align}
where $\text{Ci}$ and $\text{Si}$ are the cosine and sine integral functions respectively.

The above expression for $F_3$, obtained by assuming two independent bases, breaks down when $\mathbf{k}\parallel\mathbf{x}$ or, equivalently, $\mathbf{e}_1\parallel\mathbf{e}_2$. In general cases the expression of $\mathbf{J}$ can be shown to reduce to the following form
\begin{align}\label{eq:J_As}
\begin{split}
\mathbf{J} = 
\mathbf{A}(\mathbf {x}_{13}) e^{-i\mathbf{k}\cdot\mathbf{x}_1}
-\mathbf{A}(\mathbf {x}_{14}) e^{-i\mathbf{k}\cdot\mathbf{x}_1}
-\mathbf{A}(\mathbf {x}_{23}) e^{-i\mathbf{k}\cdot\mathbf{x}_2}
+\mathbf{A}(\mathbf {x}_{24}) e^{-i\mathbf{k}\cdot\mathbf{x}_2},
\end{split}
\end{align}

where
\begin{align}
\begin{split}
\mathbf{A}(\mathbf x) 
= &\frac{\mathbf k}{|\mathbf{k}|^2} F_1(\mathbf x) e^{i\mathbf k\cdot \mathbf x} 
+\frac{1}{|\mathbf{k}|^2}\left[-2\mathbf k  (-i \mathbf{k}\cdot \nabla) + |\mathbf{k}|^2 (-i \nabla)\right] F_2(\mathbf x)\;
\end{split}.
\end{align}
After some algebra, we arrive at the following expression
\begin{align}
\begin{split}
\mathbf{A}(\mathbf x) 
=& -\frac{\mathbf{k}}{|\mathbf{k}|^2} \left( I_0 + I_1 \right) +\frac{1}{|\mathbf{k}|^2} \left[2\mathbf{k}(\mathbf{k}\cdot\mathbf{x})-|\mathbf{k}|^2 \mathbf{x}\right] (I_2+I_3),
\end{split}
\end{align}
with
\begin{subequations}\label{eq:Is}
\begin{align}
I_0 &=\frac{e^{i\chi\rho}}{4\pi} \left[ \Ci(|\rho|\chi) -i\,\Si(\rho\chi) + \log\frac{\chi}{4|\rho|} +\gamma_E\right],\\
I_1 &=\frac{\chi}{4\pi} \int_0^1 d\alpha \, \alpha\, \frac{e^{i\alpha \rho\chi}}{\sqrt{\alpha(1-\alpha)}}
\left\{K_1 \left( \chi \sqrt{\alpha (1-\alpha)}\right) -\frac{1}{\chi \sqrt{\alpha (1-\alpha)}}\right\},\\
I_2 &= -\frac{i}{4\pi} \int_0^1 d\alpha \, e^{i\alpha\rho\chi} \left\{K_0\left( \chi \sqrt{\alpha (1-\alpha)}\right)+\log\chi\right\},\\
I_3 &= -\frac{1}{4\pi} \left( 1-e^{i\rho\chi}\right) \frac{\log\chi}{\rho\chi}.
\end{align}
\end{subequations}
Here, the $I$'s are functions of two dimensionless variables $\chi\equiv|\mathbf{k}||\mathbf{x}|$ and $\rho\equiv \cos(\phi-\theta_x)$ with  $\phi\equiv \arg \mathbf{k}$ the azimuthal angle of the gluon momentum  and $\theta_x$ the azimuthal angle of $\mathbf{x}$, such that $\mathbf{k}\cdot \mathbf{x}=\chi\rho$, and $K_n$ is the Bessel function. By numerically evaluating the above integrals, we have checked that the above two expressions for $\mathbf{J}$, eqs.~\eqref{eq:J_Fs} and \eqref{eq:J_As}, agree with each other for $\mathbf{k}\nparallel\mathbf{x}$. Moreover, they are also found to yield the same $v_n$ (within numerical errors) due to the fact that the measure for the subspace of dipole orientations corresponding to $\mathbf{k}\parallel\mathbf{x}$ is zero when one integrates over the azimuthal angles.

In the following discussions the impact parameter is taken to be aligned with the positive $x$-axis and so are the pion transverse positions during the collision:
\begin{align}\label{eq:choiceofb}
    \mathbf{x}_{A}=(b/2,0),\qquad\mathbf{x}_{B}=(-b/2,0),\qquad\text{and}\qquad\mathbf{b}=(b,0).
\end{align}
The dipole cross section for an arbitrary dipole orientation can be easily obtained by translations and rotations in the transverse plane: First, one can translate the whole collision system, moving the midpoint of its impact parameter to the origin as it is invariant under translations. Then, the dipole cross section is obtained by rotating the corresponding one for the above dipole orientation by a proper angle $\Delta\phi$ according to
\begin{align}
\left.\frac{d\hat\sigma}{d\phi}(\phi)\right|_{\text{rotated by $\Delta\phi$}}=\frac{d\hat\sigma}{d\phi}(\phi-\Delta\phi).
\end{align}
Note also that under reflections over $x$-axis and $y$-axis, it transforms respectively as
\begin{align}\label{eq:reflection}
\left.\frac{d\hat\sigma}{d\phi}(\phi)\right|_{y\to-y}=\frac{d\hat\sigma}{d\phi}(-\phi),\qquad\left.\frac{d\hat\sigma}{d\phi}(\phi)\right|_{x\to-x}=\frac{d\hat\sigma}{d\phi}(\pi-\phi)
.
\end{align}

\subsection{Momentum anisotropies in soft gluon production}

The azimuthal flow coefficients $v_n$ and their associated flow angles $\psi_n$ are defined as
\begin{align}
\frac{d\sigma}{d\phi}=\frac{\sigma}{2\pi}\left[1+2 \sum\limits_{n} v_n \cos(n(\phi-\psi_n))\right]
\;,
\end{align}
such that $v_n>0$\footnote{
As discussed in the following sections, $\psi_n$ is not necessarily equal to the reaction plane angle in the corresponding classical collision geometry.
}. Here, the dependence of $\sigma$ on $\mathbf{b}, \eta$ and $k_T$ are omitted for brevity. To facilitate our discussions below, we define two-dimensional flow coefficients $\mathbf{v}_n$ with their components respectively given by
\begin{align}\label{eq:vn_def}
    v_n^x =\frac{1}{\sigma} \int_0^{2\pi}{d\phi}\frac{d\sigma}{d\phi}\cos(n\phi),\qquad
    v^y_n =\frac{1}{\sigma} \int_0^{2\pi}{d\phi}\frac{d\sigma}{d\phi}\sin(n\phi).
\end{align}
That is,
\begin{align}
\frac{d\sigma}{d\phi}=\frac{\sigma}{2\pi}\left[1+2 \sum\limits_{n} (v_n^x\cos(n\phi)+v_n^y\sin(n\phi))\right]
\;.
\end{align}
And one has, accordingly,
\begin{align}\label{eq:vnpsin}
    v_n=|\mathbf{v}_n|,\qquad \psi_n=\frac{1}{n}\phi_{\mathbf{v}_n}
\end{align}
with $\phi_{\mathbf{v}_n}$ the azimuthal angle of $\mathbf{v}_n$ around the $x$-axis.

\section{Momentum anisotropies in dipole-dipole scattering}\label{sec:dipole_dipole_v2}
In this section, we study transverse momentum anisotropies, mainly $v_2$, of soft gluon production in dipole-dipole scattering. We first study the $k_T$-dependence of $v_2$ and evaluate it analytically both at low and high $k_T$. Then, we calculate $v_2$ for the intermediate $k_T$ regime numerically with the afore-derived exact expression and investigate the correlation between momentum anisotropies and dipole orientations.

\subsection{The low $k_T$ limit}
In the limit $k_T \to 0$ when the de Broglie wavelength of the produced gluon is larger than the dipole sizes and the impact parameter, momentum anisotropies are expected to vanish. This can be explicitly checked from eq.~\eqref{eq:sigppg}:
 \begin{align}
     \frac{d\hat\sigma}{d^2\mathbf{b} d\eta d^2 \mathbf{k} }\to\frac{\alpha_s N_c}{\pi^2}\frac{1}{|\mathbf{k}|^2}\frac{d\hat{\sigma}}{d^2\mathbf{b}}\qquad\text{as $k_T\to0$},
\end{align}
which shows explicitly $v_n\to0$ as $k_T\to0$. Here, the LO total dipole cross section is given in eq.~\eqref{eq:sigqqb}.

Let us evaluate the behavior of $\mathbf{v}_2$ at low $k_T$ via the expressions of $|\mathbf J|^2$.  
The small $k_T$ limit of $|\mathbf J|^2$ can be obtained by directly expanding its components $I_0$, $I_1$, $I_2$ and $I_3$ [see eq.~\eqref{eq:Is}] at $\chi=0$; recall that $\chi=|\mathbf{k}||\mathbf{x}|$. 
The angular dependence resides in $\rho[=\cos(\theta_x-\phi)]$, and here we keep $|\mathbf J|^2$ up to $\mathcal{O}(\chi^0)$, the lowest order in $\chi$ containing $\rho$.
The corresponding expansions of $I$'s are
\begin{subequations}
\begin{align}
\begin{split}
&\lim_{\chi\to 0} I_0 = \frac{1}{4\pi}\left(\log\frac{\chi^2}{4}+2\gamma_E\right)
+\frac{i\rho\chi}{4\pi}\left(\log\frac{\chi^2}{4}+2\gamma_E-1\right)
\\
&\qquad 
-\frac{\rho^2\chi^2}{8\pi}\left(\log\frac{\chi^2}{4}+2\gamma_E-\frac{3}{2}\right)
+\mathcal{O}(\chi^3),
\end{split}\\
&\lim_{\chi\to 0} I_1 = \frac{\chi^2}{32\pi}\left(\log\frac{\chi^2}{4}+2\gamma_E-3\right)+\mathcal{O}(\chi^3),\\
&\lim_{\chi\to 0} I_2+I_3 =-\frac{1}{16\pi}\left(-2i+\rho\chi\right)\left(\log\frac{\chi^2}{4}+2\gamma_E-2\right)
+\mathcal{O}(\chi^2)
.
\end{align}
\end{subequations}
The expansion for $\mathbf{A}(\mathbf{x})$ follows as
\begin{align}
\begin{split}
\lim_{k_T \to 0}  \mathbf{A}(\mathbf{x}) =
& 
-\frac{\mathbf {k}}{|\mathbf {k}|^2}
 \frac{1}{4\pi}\left(\log\frac{|\mathbf {k}|^2|\mathbf {x}|^2}{4}+2\gamma_E\right)
 +\frac{\mathbf x}{16\pi} (\mathbf {k}\cdot \mathbf {x})\left(\log\frac{|\mathbf {k}|^2 |\mathbf {x}|^2}{4}+2\gamma_E-2\right)\\
&+\frac{\mathbf {k}}{|\mathbf {k}|^2}
\left[
\frac{(\mathbf {k}\cdot\mathbf {x})^2}{16\pi}
-\frac{\mathbf {k}^2\mathbf {x}^2}{32\pi}\left(\log\frac{\mathbf {k}^2\mathbf {x}^2}{4}+2\gamma_E-3\right)\right]
\\
& -i\frac{\mathbf {k}}{|\mathbf {k}|^2}
\frac{(\mathbf {k}\cdot\mathbf {x})}{4\pi}
-i\frac{ \mathbf  x}{8\pi} \left(\log\frac{\mathbf {k}^2|\mathbf {x}|^2}{4}+2\gamma_E-2\right) +\mathcal{O}(|\mathbf {k}|^2).
\end{split}
\end{align}

The resulting $|\mathbf{J}|^2$ has the following format
\begin{align}
\begin{split}
\lim_{k_T \to 0}|\mathbf J|^2=&\frac{1}{|\mathbf k|^2} 
\Bigg[
B_0 + |\mathbf k|^2 \sum_{i,j=1}^4 D_{ij} \cos(\theta_i-\phi)\cos(\theta_j-\phi)
\Bigg]
+\mathcal{O}(|\mathbf k|)
\;,
\end{split}
\end{align}
in which $\theta_i=\text{arg}~\mathbf{x}_i$ and the coefficients $B_0$ and $D_{ij}$ do not depend on the angle $\phi(=\arg \mathbf k)$:
\begin{align*}
& B_0=\frac{1}{16\pi^2 } \left(\log\frac{|\mathbf {x}_{13}|^2 |\mathbf {x}_{24}|^2}{|\mathbf {x}_{14}|^2|\mathbf {x}_{23}|^2}\right)^2\;,\qquad
  D_{ij}=-\frac{1}{32 \pi^2 } \log\frac{|\mathbf {x}_{13}|^2 |\mathbf {x}_{24}|^2}{|\mathbf {x}_{14}|^2|\mathbf {x}_{23}|^2}|\mathbf x_i||\mathbf x_j| f_{ij},\\
&
f_{ij}=\begin{cases}
(-1)^{i+1}\log\dfrac{|\mathbf x_{i,l(i)}|^2}{|\mathbf x_{i,l(i)+1}|^2}, &\text{if }  i=j\\
(-1)^{i-j}\left[1-\log\dfrac{|\mathbf k|^2 |\mathbf x_{ij}|^2}{4}\right], &\text{if }  i\neq j
\end{cases}
\;,\qquad l(i)\equiv\begin{cases}3, &\text{if } i=1,2\\
1, &\text{if } i=3,4
\end{cases}.
\end{align*}
Integrating over $\phi$, one obtains
\begin{subequations}
\begin{align}
&\lim_{k_T \to 0}\int_0^{2\pi} \diff\phi |\mathbf J|^2 = \frac{2\pi}{|\mathbf k|^2}\left[B_0+\frac{|\mathbf k|^2}{2}\sum_{i,j=1}^4 D_{ij} \cos(\theta_i-\theta_j) \right]+\mathcal{O}(|\mathbf k|),\\
&\lim_{k_T \to 0}\int_0^{2\pi} \diff\phi |\mathbf J|^2 \cos(2\phi) =
\frac{\pi}{2} \sum_{i,j=1}^4  D_{ij} \cos(\theta_i+\theta_j)+\mathcal{O}(|\mathbf k|).
\end{align}
\end{subequations}
Note that here we are evaluating $v_2^x$ as defined in eq.~\eqref{eq:vn_def}, and the calculation of $v_2^y$ is similar. Besides, for the reason that will be explained later, $v_2^y$ vanishes for azimuthally-integrated dipoles (i.e., as in the pion-pion case) and thus $v_2=|v_2^x|$.
It follows that $v_2$ scales as $k_T^2$ at small $k_T$:
\begin{align}\label{eq:v2lowkT}
\lim_{k_T \to 0} v_2=\frac{\sum_{i,j=1}^4 D_{ij} }{4 B_0}\cos(\theta_i+\theta_j) |\mathbf k|^2 +\mathcal{O}(|\mathbf k|^3 )
\;.
\end{align}

\subsection{The high $k_T$ limit}
At high $k_T$ when the gluon has a de Broglie wavelength much shorter than the dipole sizes and the impact parameter, one may expect that the azimuthal distribution of the gluon is only sensitive to its emitter ($q$ or $\bar{q}$) and, hence, isotropic. This is also needed in order to justify that the production of high-$k_T$ partons
can be sufficiently described by the single parton distribution functions or the thickness beam functions of pions, which are rotationally symmetric in the transverse plane~\cite{Wu:2021ril}.

In order to evaluate the behavior of $\mathbf{v}_2$ at large $k_T$ via $|\mathbf J|^2$, we use the expressions of $F_1$ and $F_2$ and expand $|\mathbf J|^2 $ at $1/|\mathbf k|=0$.
We apply the method of regions on $F_2$, by noting two dominant regions: $\mathbf{l}\sim \mathbf{k}$ and $\mathbf{l}\sim \mathbf 0$. These two regions are well-separated in the large $k_T$ limit, and one obtains
\begin{align}
\begin{split}
\lim_{k_T \to \infty}F_2(\mathbf x) 
=&
\mu^{2\epsilon}
\int\frac{d^d l}{(2\pi)^d}
\frac{e^{i\mathbf{l}\cdot \mathbf{x}}}{|\mathbf l|^2}
      \left[
      \frac{1}{|\mathbf k|^2}+\frac{2 \mathbf k\cdot\mathbf l}{|\mathbf k|^4} 
      \right]
+
\mu^{2\epsilon}
\int\frac{d^d l}{(2\pi)^d}
\frac{e^{i\mathbf{l}\cdot \mathbf{x}}}{|\mathbf{l}-\mathbf{k}|^2}
    \left[
    \frac{3}{|\mathbf k|^2}-\frac{2 \mathbf k\cdot \mathbf l}{|\mathbf k|^4}
     \right]\\
     &
+\mathcal{O}\left(\frac{1}{|\mathbf{k}|^4}\right)
\\
=
-& \frac{1+e^{i\mathbf{k}\cdot \mathbf{x}}}{4\pi|\mathbf k|^2}\left[ \frac{1}{\epsilon} + \log(|\mathbf x|^2\mu^2) +\gamma_E +\log\pi \right]
+\frac{1-e^{i\mathbf{k}\cdot \mathbf{x}}}{\pi|\mathbf k|^4}\frac{i\mathbf k\cdot \mathbf x}{|\mathbf x|^2}
+\mathcal{O}\left(\frac{1}{|\mathbf{k}|^4}\right),
\end{split}
\end{align}
where we have applied the expression of $F_1$ in eq.~\eqref{eq:F1}. The expansion for $\mathbf{A}(\mathbf{x})$ follows as
\begin{align}
\lim_{k_T \to \infty}  \mathbf{A}(\mathbf{x}) =
-\frac{i \mathbf k}{\pi |\mathbf k|^4 |\mathbf x|^2} ( \mathbf{k}\cdot\mathbf{x}) 
+\frac{ i \mathbf x}{2\pi |\mathbf k|^2 |\mathbf x|^2}(1+e^{i\mathbf{k}\cdot\mathbf{x}}) 
+\mathcal{O}\left(\frac{1}{|\mathbf{k}|^3}\right)
\;.
\end{align}
The resulting $|\mathbf{J}|^2$ is in the following format
\begin{align}
\begin{split}
\lim_{k_T \to \infty}|\mathbf{J}|^2=&
\frac{1}{|\mathbf k|^4} 
\Bigg\{
B_1 + \bigg[
\sum_{\substack{m,n=1\\ m\neq n}}^4  f_{mn}
+
\sum_{\substack{n=1,2 \\ m=3,4}}
g_{mn}\cos(2\phi)
+\sum_{\substack{n=1,2 \\ m=3,4}} h_{mn}\sin(2\phi)
\bigg] \\
& \cos\big[|\mathbf k| |\mathbf x_m|\cos(\theta_m-\phi)-|\mathbf k| |\mathbf x_n|\cos(\theta_n-\phi)\big]
 \Bigg\}
+\mathcal{O}\left(\frac{1}{|\mathbf{k}|^6}\right)
\;,
\end{split}
\end{align}
in which the coefficients $B_1, f_{mn}, g_{mn}, h_{mn}$ do not depend on $\phi$:
\begin{align*}
& \mathbf Q_1\equiv 
    -\frac{\mathbf x_{13}}{x_{13}^2}
    +\frac{\mathbf x_{14}}{x_{14}^2}
\;,
\qquad 
\mathbf Q_2\equiv 
    \frac{\mathbf x_{23}}{x_{23}^2}
    -\frac{\mathbf x_{24}}{x_{24}^2}
\;,\qquad
\mathbf Q_3\equiv 
    -\frac{\mathbf x_{13}}{x_{13}^2}
    +\frac{\mathbf x_{23}}{x_{23}^2}
\;,
\qquad 
\mathbf Q_4\equiv 
    \frac{\mathbf x_{14}}{x_{14}^2}
    -\frac{\mathbf x_{24}}{x_{24}^2}
\;,\\
& B_1\equiv \frac{1}{4\pi^2 } 
\sum_{i=1}^4 |\mathbf Q_i|^2
,\qquad
 f_{mn}\equiv \frac{1}{4\pi^2 } \mathbf Q_m\cdot \mathbf Q_n,
 \qquad \theta_{Q_n}=\arg \mathbf Q_n,\\
& g_{mn}\equiv -\frac{1}{2\pi^2 } | \mathbf Q_m| | \mathbf Q_n|\cos(\theta_{Q_m}+\theta_{Q_n}),\qquad
 h_{mn}\equiv -\frac{1}{2\pi^2 } | \mathbf Q_m| | \mathbf Q_n|\sin(\theta_{Q_m}+\theta_{Q_n})\;.
\end{align*}

Integrating over $\phi$, one obtains
\begin{subequations}
\begin{align}
\begin{split}
   & \lim_{k_T \to \infty}\int_0^{2\pi} \diff \phi |\mathbf{J}|^2 =
\frac{2\pi}{ |\mathbf k|^4}
B_1
+\frac{2\pi}{ |\mathbf k|^4}
\left[ \sum_{\substack{m,n=1\\ m\neq n}}^4  f_{mn}J_0[|\mathbf k|  \bar x_{mn}]
-\sum_{\substack{n=1,2 \\ m=3,4}} g_{mn}J_2[|\mathbf k|  \bar x_{mn}]
\right]\\
&\qquad
+\mathcal{O}\left(\frac{1}{|\mathbf{k}|^6 }\right)\;,
\end{split}\\
\begin{split}
&\lim_{k_T \to \infty} \int_0^{2\pi} \diff \phi |\mathbf{J}|^2 \cos(2\phi) = \frac{2\pi}{ |\mathbf k|^4}
\left[
 - \sum_{\substack{m,n=1\\  m\neq n}}^4 f_{mn}J_2[|\mathbf{k}| \bar x_{mn}]
+\sum_{\substack{n=1,2 \\ m=3,4}} g_{mn}J_0[|\mathbf{k}| \bar x_{mn}]
\right]\\
&\qquad
+\mathcal{O}\left(\frac{1}{|\mathbf{k}|^5}\right)\;,
\end{split}
\end{align}
\end{subequations}
in which $\bar x_{mn}\equiv x_m^x+x_m^y-x_n^x-x_n^y$, and $J_t$ is the $t$-th Bessel J function. It follows that $v_2$ scales as $1/\sqrt{k_T}$ at large $k_T$:
\begin{align}\label{eq:v2_large_kt}
\begin{split}
\lim_{k_T \to \infty}  v_2
=&\frac{
\left[
 - \sum_{\substack{m,n=1\\  m\neq n}}^4 f_{mn}J_2[|\mathbf k|  \bar x_{mn}]
+\sum_{\substack{n=1,2 \\ m=3,4}}g_{mn}J_0[|\mathbf k|  \bar x_{mn}]
\right]
+\mathcal{O}\left(1/|\mathbf k|\right)
}{
B_1+
\left[  \sum_{\substack{m,n=1\\  m\neq n}}^4 
 f_{mn}J_0[|\mathbf k|  \bar x_{mn}]
-\sum_{\substack{n=1,2 \\ m=3,4}}g_{mn}J_2[|\mathbf k|  \bar x_{mn}]
\right]
+\mathcal{O}\left(1/|\mathbf k|^2\right)}\\
=
\frac{1}{B_1}&
\sqrt{\frac{2}{\pi  |\mathbf k| }}
 \left[
 \sum_{\substack{m,n=1\\  m\neq n}}^4 \frac{f_{mn}}{\sqrt{ \bar x_{mn}} }\cos(|\mathbf k|  \bar x_{mn}-\frac{\pi}{4})
+\sum_{\substack{n=1,2 \\ m=3,4}}\frac{g_{mn}}{\sqrt{ \bar x_{mn}}}\cos(|\mathbf k|  \bar x_{mn}-\frac{\pi}{4})
\right]\\
&
+\mathcal{O}(\frac{1}{|\mathbf k| })
\;.
\end{split}
\end{align}
Strictly speaking, this result is valid at large $k_T\bar x_{mn}$ instead of large $k_T$. One can see that the large-$k_T$ expression of $|\mathbf J|^2$ diverges at $\bar x_{mn}=0$ (e.g., $x_{13}=0 $), whereas the full expression in sec.~\ref{eq:eval_sig_dipole} is finite. 

\subsection{The intermediate $k_T$-dependence}
\label{sec:v2kTdipole}

In this subsection we study the transition of $v_2$ from the quadratic growth at low $k_T$ to the $1/\sqrt{k_T}$ suppression at high $k_T$. For such an intermediate range of $k_T$, we use the full result of the dipole cross section in sec.~\ref{eq:eval_sig_dipole} to evaluate $v_2$ numerically. Since $v_2$ in pion-pion collisions results from the superposition of gluons emitted off all the possible dipole orientations weighted by the pion wave functions, below we investigate the dependence of $v_2$ on dipole orientations, characterized by $b$, $r_A, \phi_A, r_B$ and $\phi_B$ with $r_i$ and $\phi_i$ respectively the azimuthal angle and the modulus of $\mathbf{r}_i$. As motivated by the fact that the valence $q$ or $\bar{q}$ typically carries half of the ``+" momentum of its parent pion, we take $\xi_A=\xi_B=1/2$ in the following discussions.

\subsubsection{$v_2$ for ``typical" dipole orientations with $\phi_A=\phi_B=\pi/2$
}

As shown in the following sections, some main qualitative features of $v_2$ for the dipole configurations with $(\phi_A,\phi_B)=(\pi/2,\pi/2)$ survive the integration over the LFWFs in pion-pion collisions. In order to discuss such qualitative features, let us first fix $b$ (along the $x$-axis according to eq.~(\ref{eq:choiceofb})) and vary the dipole sizes $r_A=r_B\equiv\;d_{dp}$ with both dipole azimuthal angles fixed to be $\pi/2$.  For such dipole configurations one has $\frac{d\sigma}{d\phi}(\phi)=\frac{d\sigma}{d\phi}(-\phi)$ according to eq.~\eqref{eq:reflection} and, as a result, $v_2^y$ vanishes.

\begin{figure*}[htp!]
    \centering
    \includegraphics[width=0.6\textwidth]{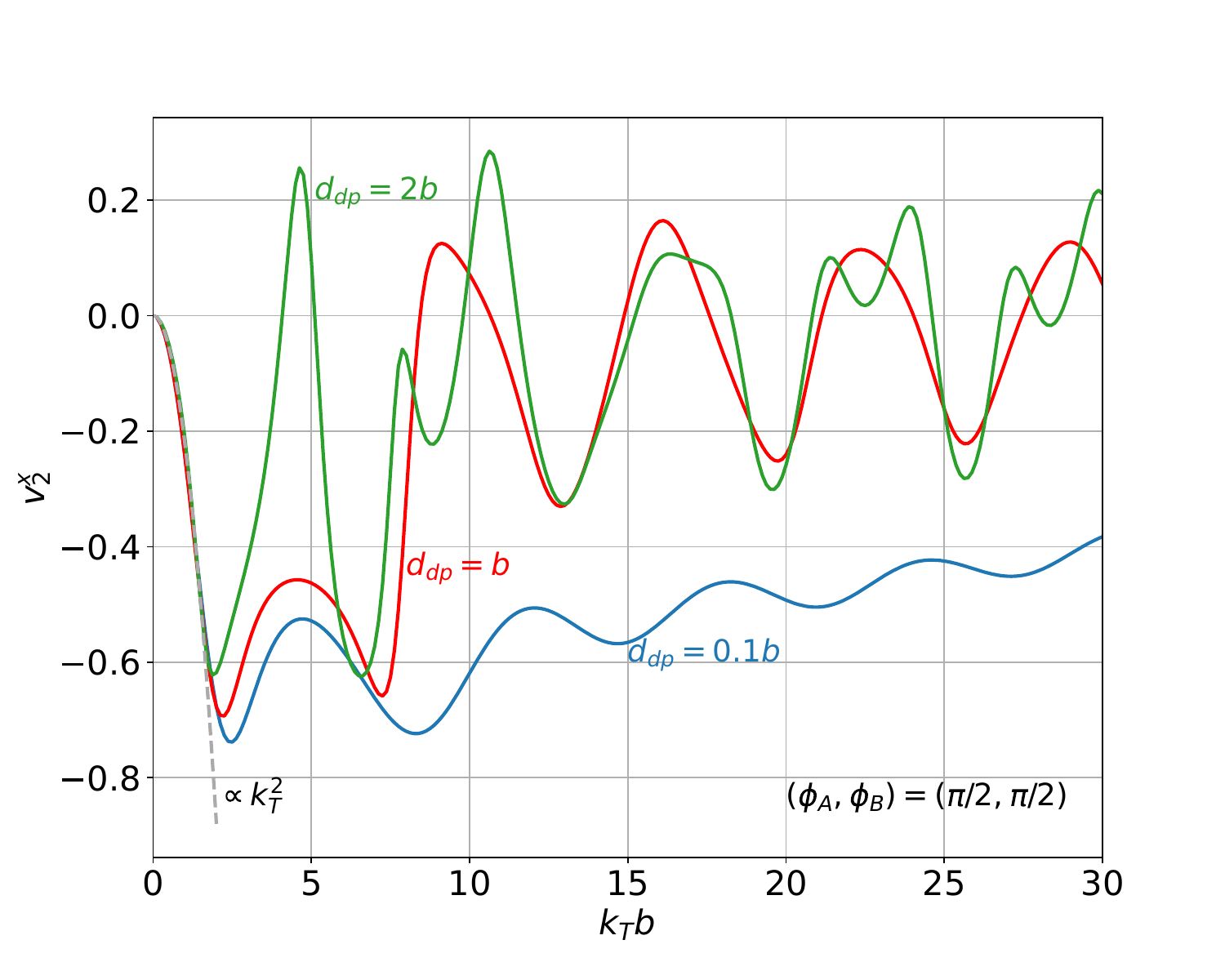}
    \caption{$v^x_2$ as a function of $k_T$ at $b$ for different dipole sizes with $(\phi_A,\phi_B)=(\pi/2,\pi/2)$. Here, both dipoles have the same size $d_{dp}$.}
    \label{fig:v2b1rdp}
\end{figure*}

Figure~\ref{fig:v2b1rdp} shows the dependence of $v_2^x$ on $k_T$ for different dipole sizes: $d_{dp}=0.1b, b$ and $2b$. For all these dipole sizes, $v_2^x$ is observed to grow quadratically in $k_T$ at small $k_T$ while it damps and oscillates at large $k_T$.
In the shown range of $k_T\leq 30b^{-1}$, the curves for $d_{dp}=b$ and $2b$ can be fairly approximated by our analytic results both at low and high $k_T$ while the curve for $d_{db}=0.1b$ approaches to the asymptotic high-$k_T$ behavior only at $k_T>30b^{-1}$ when $k_T \bar x_{mn}$, as in eq.~\eqref{eq:v2_large_kt}, is sufficiently large.

At intermediate $k_T$, Fig.~\ref{fig:v2b1rdp} shows a qualitative difference between the more central ($d_{dp}=2b$) and the more peripheral ($ d_{dp}=b$ and $0.1b$) collisions. At $k_T \sim 1/b$, all the curves of $v_2=|v_2^x|$ start to deviate from the quadratic growth and develop their first peak (corresponding to the first minimum of $v_2^x$). Then, as $k_T$ increases, $v_2^x$ for $d_{dp}=2b$ changes its sign after passing its first minimum. In contrast, there is no sign change in the curves for $d_{dp}=b$ or $d_{dp}=0.1b$ before they develop their second minimum; and no sign change is observed for $d_{dp}=0.1b$ in the shown range of $k_T$ up to $30b^{-1}$. Such a sign change after the first minimum of $v_2^x$ is found to exist for $d_{dp}\gtrsim\;1.5b$.

\subsubsection{$v_2$ for different dipole azimuthal angles}

Let us further investigate the dependence of $v_2$ on $(\phi_A, \phi_B)$. The dipole sizes $r_A$ and $r_B$, denoted by $d_{dp}$, are now kept fixed (it could be estimated as the hadron size or the charge radius defined in the next section). As we are only interested in collisions dominated by the strong interaction, below we investigate in detail two sets of dipole orientations with $b=0.5 d_{dp}$ and $b=d_{dp}$ respectively.

\begin{figure*}[htp!]
    \centering
    \includegraphics[width=0.6\textwidth]{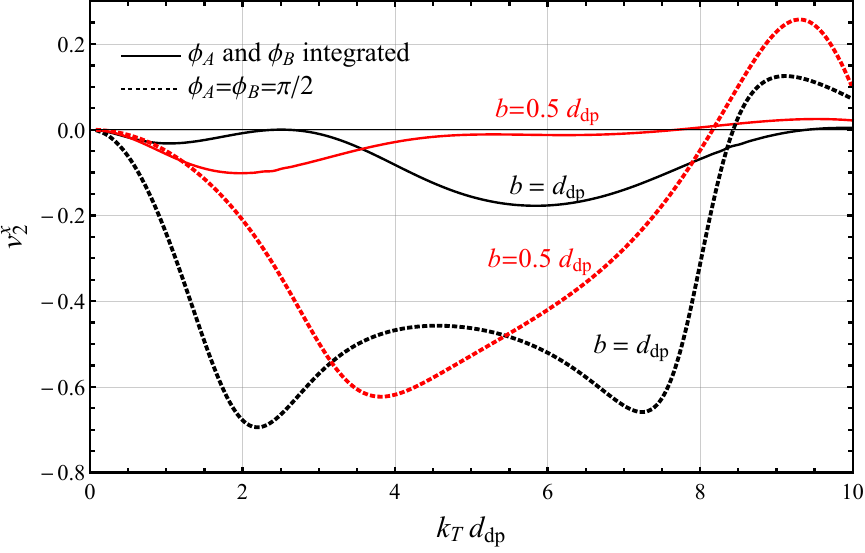}
    \caption{The $(\phi_A,\phi_B)$-integrated $v^x_2$ as a function of $k_T$ at $b=0.5d_{dp}$ and $b=d_{dp}$. To facilitate comparison, the results (dashed) for $(\phi_A,\phi_B)=(\pi/2,\pi/2)$ in Fig.~\ref{fig:v2b1rdp} are replotted as a function of $k_T d_{dp}$ in this figure.}
    \label{fig:v2b}
\end{figure*}

Since the pion wave function squared respects rotational symmetry (in the transverse plane), we first study the net effect of integrating over $\phi_A$ and $\phi_B$ on $v_2$. In this case the contributions to $v_n^y$ from $(\phi_A,\phi_B)$ and $(-\phi_A,-\phi_B)$ cancel due to reflection symmetry over the $x$-axis and, accrodingly, the $(\phi_A,\phi_B)$-integrated $v_n^y=0$. The results of $v_2^x$ for $b=0.5 d_{dp}$ and $b=d_{dp}$, in comparison with those for the typical dipole orientations with $(\phi_A,\phi_B)=(\pi/2,\pi/2)$, are shown in Fig.~\ref{fig:v2b}. The main qualitative features observed for $(\phi_A,\phi_B)=(\pi/2,\pi/2)$ survive: the curve for $b=0.5 d_{dp}$ only develops one (global) minimum while the other one for $b= d_{dp}$ has two minima before entering the suppression region at high $k_T$. Quantitatively, the absolute values of the minima are smaller and their locations are shifted to lower $k_T$ compared to those for  $(\phi_A,\phi_B)=(\pi/2,\pi/2)$. Moreover, the second minimum of the $b= d_{dp}$ curve becomes the global minimum unlike that for $(\phi_A,\phi_B)=(\pi/2,\pi/2)$.

\begin{figure*}[htp!]
    \centering
    \includegraphics[width=0.48\textwidth]{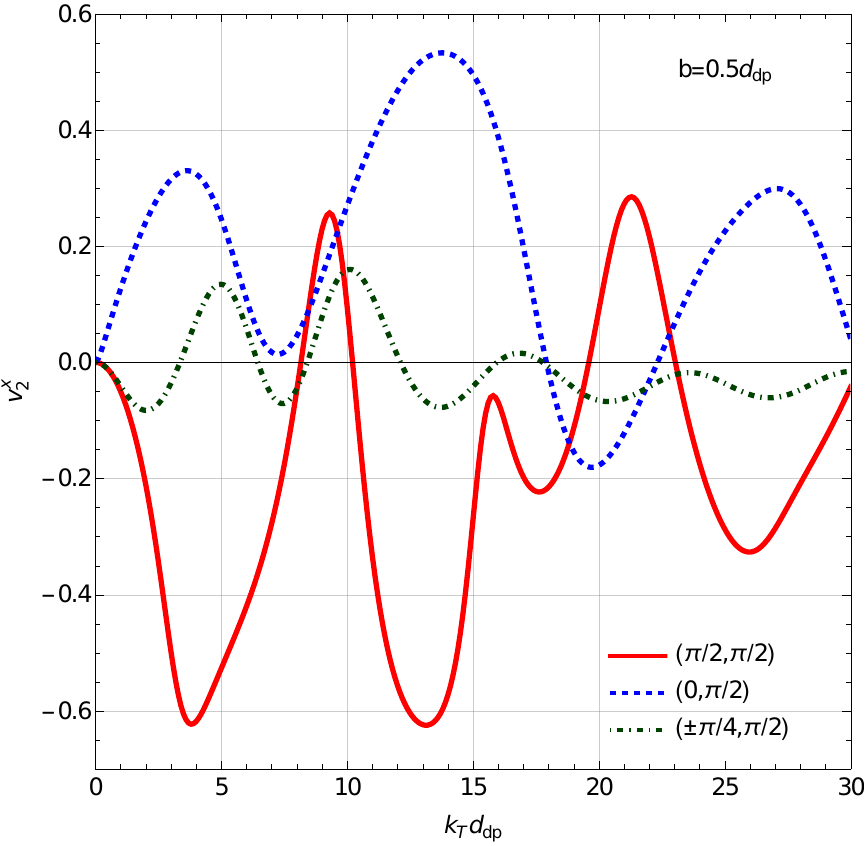}
    \includegraphics[width=0.48\textwidth]{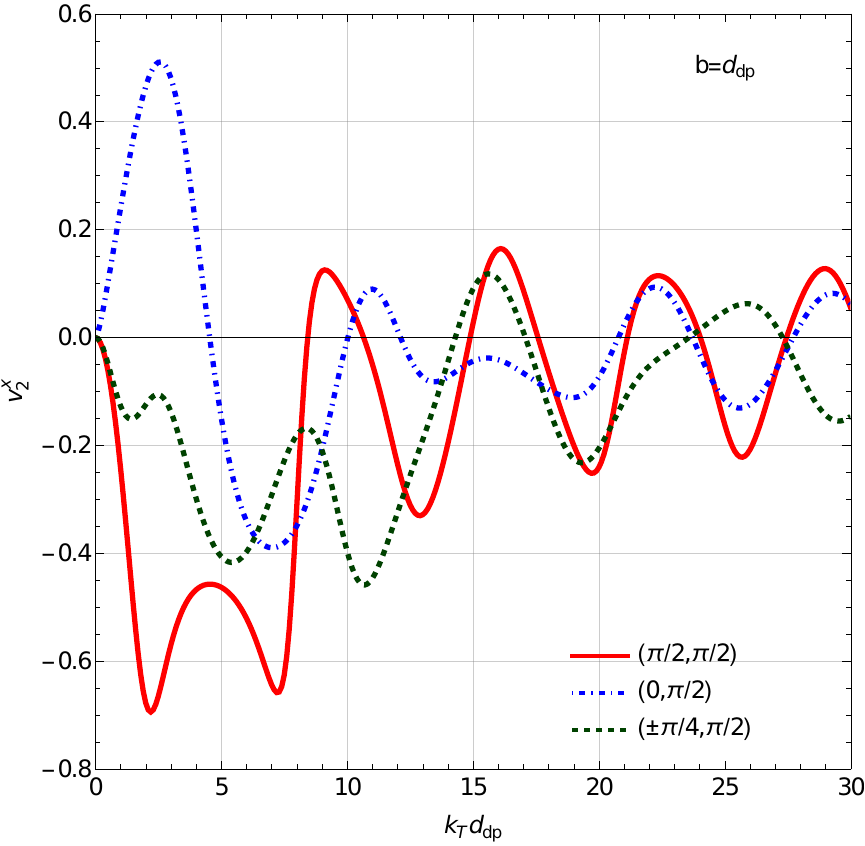}
    \caption{$v^x_2$ as a function of $k_T$ for different values of $(\phi_A, \phi_B)$ at $b=0.5d_{dp}$ (left) and $b=d_{dp}$ (right).}
    \label{fig:v2b051}
\end{figure*}

In order to understand in detail how the superposition of gluon waves produced by different dipole orientations is responsible for the above $(\phi_A,\phi_B)$-integrated results, we show $v_2^x$ for $(\phi_A,\phi_B)=(0,\pi/2)$ and $(\pm\pi/4,\pi/2)$ in comparison with $(\phi_A,\phi_B)=(\pi/2,\pi/2)$ respectively for $b=0.5d_{dp}$ (left) and $b=d_{dp}$ (right) in Fig.~\ref{fig:v2b051}. 

As shown in both plots of Fig.~\ref{fig:v2b051}, at low $k_T$ (with $k_Td_{dp} < 3$) only the $(0,\pi/2)$ curve is positive, which peaks at about the same value of $k_T$ as the first minimum of the corresponding $(\pi/2,\pi/2)$ curve. That is, around this region it is superposed destructively with the $(\pi/2,\pi/2)$ curve in its contribution to the $(\phi_A,\phi_B)$-integrated $v_2^x$.  And this is consistent with the fact that the minimum in the $(\pi/2, \pi/2)$ curve around this region is destroyed and the $(\phi_A,\phi_B)$-integrated $v_2^x$ instead develops its first minimum at lower $k_T$ for both values of $b$ as observed in Fig.~\ref{fig:v2b}.

At higher $k_T$ the $(\pi/2,\pi/2)$ curve for $b=0.5 d_{dp}$ (in the left plot of Fig.~\ref{fig:v2b051}) changes its sign and develops a maximum while the one for $b=d_{dp}$ (in the right plot of Fig.~\ref{fig:v2b051}) never changes its sign before developing another minimum around $k_T=7.5 d_{dp}^{-1}$, as first shown in Fig.~\ref{fig:v2b1rdp}. All the other curves for $b=d_{dp}$ are negative around the location of the second minimum of the $(\pi/2,\pi/2)$ curve. Accordingly, the gluon waves from these dipole orientations are all superposed constructively in their contributions to the $(\phi_A,\phi_B)$-integrated $v_2^x$.  In contrast, for $b=0.5 d_{dp}$ one can not convincingly identify a higher $k_T$ region dominated by such constructive superposition.

\subsubsection{The azimuthal distribution for different dipole orientations}\label{sec:vndipole}
At the end, we study the azimuthal distribution defined as a two-dimensional vector: $\frac{1}{\hat\sigma}\frac{d\hat\sigma}{d\phi}(\cos\phi,\sin\phi)$. For brevity $\hat\sigma$ here denotes the $\phi$-integrated differential dipole cross section at given $\eta$, $k_T$ and $b$, according to eq.~\eqref{eq:sigppg}. The magnitude of this vector represents the probability density for the gluon to be emitted at an angle $\phi$ with respect to the $x$-axis. And in the azimuthal distribution plot the flow angle $\psi_2=\psi_{\mathbf{v}_2}/2$ can be intuitively estimated as the angle of the long-axis of an ellipse (or a ``spindle") tightly enclosing the distribution curve.

\begin{figure*}[htp!]
    \centering
    \includegraphics[width=\textwidth]{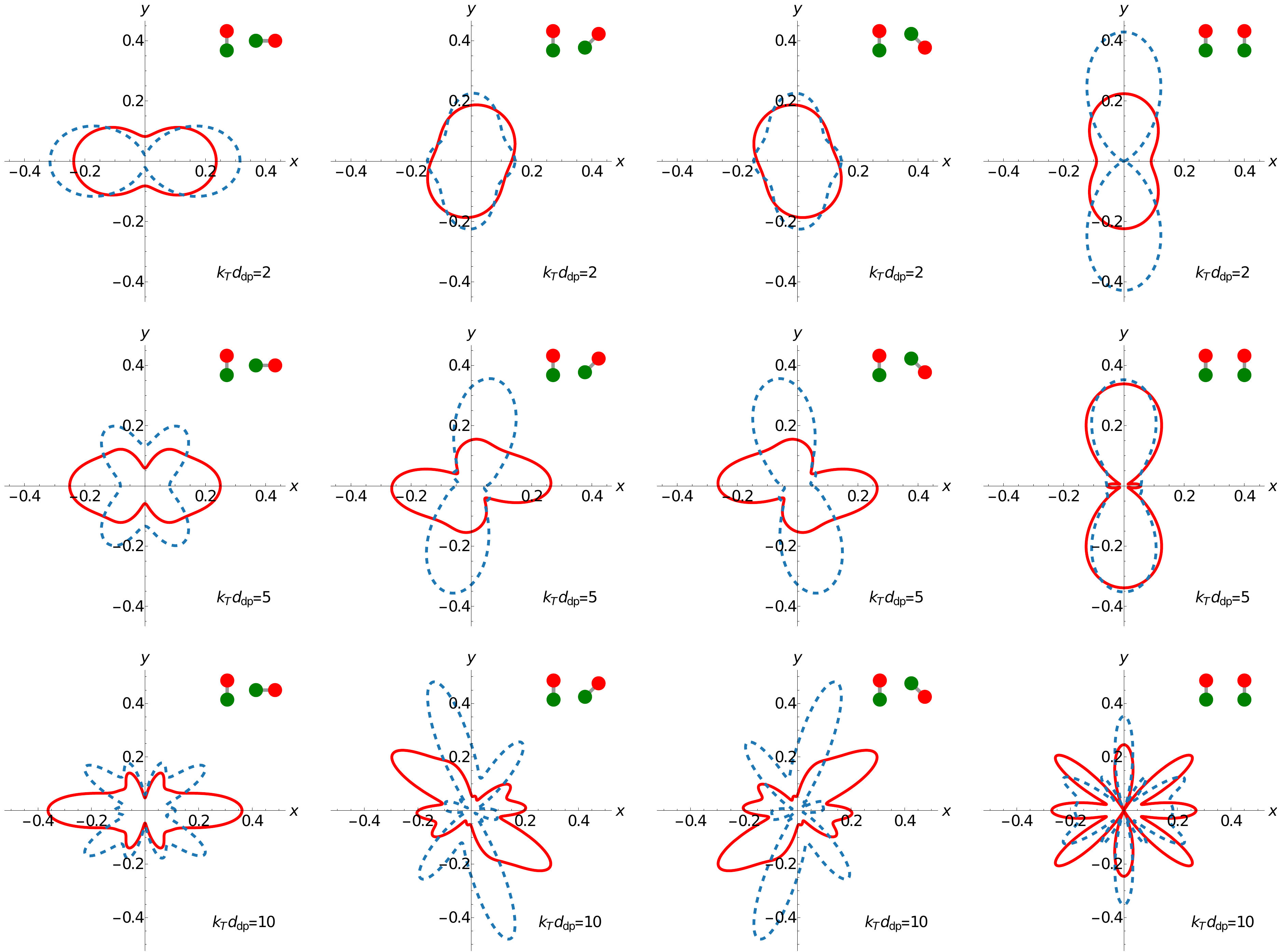}\\
    \caption{The azimuthal distribution $\frac{1}{\hat\sigma}\frac{d\hat\sigma}{d\phi}(\cos\phi,\sin\phi)$ at $b=0.5 d_{dp}$ (red solid) and $b= d_{dp}$ (blue dashed) for $k_T d_{dp}=2,5$ and $10$. The dipole orientations are illustrated in the top right corner of the plots.}
    \label{fig:azimuthalddp}
\end{figure*}

The azimuthal distributions at $k_T=2 d_{dp}^{-1}, 5d_{dp}^{-1}$ and $10d_{dp}^{-1}$ for the same dipole orientations as those in Fig.~\ref{fig:v2b051} are shown in Fig.~\ref{fig:azimuthalddp}. At $k_T=2 d_{dp}^{-1}$ (the first row in this figure), the distributions for both $b=0.5 d_{dp}$ (red solid) and $b=d_{dp}$ (blue dashed) are qualitatively very similar for all the shown 4 dipole orientations: Only for $(\phi_A,\phi_B)=(0,\pi/2)$ the gluon is most probably emitted along the $x$-axis, so $\psi_2 = 0$ and $\phi_{\mathbf{v}_2} = 0$, corresponding to a positive $v_2^x$. For the other 3 dipole orientations the most probable direction to emit a gluon is close to the $y$-axis, corresponding to a negative $v_2^x$.

The plots in the second row of Fig.~\ref{fig:azimuthalddp} show the azimuthal distributions at $k_T=5 d_{dp}^{-1}$. Only for $(\phi_A,\phi_B)=(\pi/2,\pi/2)$, $v_2^x$ remains negative for both values of $b$. For $(\phi_A,\phi_B)=(0,\pi/2)$, it is evident that the most probable emission still occurs along the $x$-axis for $b=0.5 d_{dp}$ while it is, after eliminating higher harmonics, along the $y$-axis for $b=d_{dp}$ (that is, its $v_2^x$ becomes negative). For $(\phi_A,\phi_B)=(\pm\pi/4,\pi/2)$, $v_2^x$ at $b=0.5 d_{dp}$ changes its sign and becomes positive while $v_2^x$ at $b=d_{dp}$ remains negative. As a result, the distributions from the 4 dipole orientations are superposed constructively for $b=d_{dp}$ but destructively for $b=0.5d_{dp}$ when $(\phi_A,\phi_B)$ are integrated over.

The azimuthal distributions at $k_T=10 d_{dp}^{-1}$ are shown in the last row of Fig.~\ref{fig:azimuthalddp}. For both values of $b$, the distributions become more oscillatory than low $k_T$ (it becomes even more so at higher $k_T$). The values of $v_2^x$ at $b=0.5 d_{dp}$ are positive for all the four dipole orientations, and $v_2^x$ at $b= d_{dp}$ is negative for $(\phi_A,\phi_B)=(\pm\pi/4,\pi/2)$ and close to 0 for the other two orientations, cf. Fig.~\ref{fig:v2b051}. Superposing all the highly oscillatory distributions in the $(\phi_A,\phi_B)$ space, as illustrated by these dipole orientations, tends to isotropize the gluon emission. This is consistent with the fact that the single-parton distributions (thickness beam functions), used to describe hard processes, are isotropic. Therefore, such a highly oscillatory behavior at high $k_T$ could be viewed as an evidence for the validity of QCD factorization.

\section{Momentum anisotropies in pion-pion collisions}\label{sec:pion_pion}
In this section we study momentum anisotropies in pion-pion collisions by convoluting the dipole cross section with pion wave functions. Prior to discussing our main results, we first briefly review the LFWFs used in our calculations.

\subsection{The light-front wave function}
\label{sec:wavefunction}
From a previous work~\cite{Qian:2020utg}, we obtain the LFWF for the low-lying states in the light meson system. Within the valence $\ket{q \bar{q}}$ Fock sector, the state vector of a meson $h$ reads
\begin{align}\label{eq:valence_WF}
  \ket{\psi_h(P, j, m_j)} 
  = &\sum_{s_q, s_{\bar{q}}}\int_0^1\frac{\text{d}\xi}{2\xi(1-\xi)}\int\frac{\text{d}^2\mathbf{k}}{(2\pi)^3}\psi_{s_q, s_{\bar{q}}/h}^{(m_j)}(\mathbf{k}, \xi) 
  \nonumber \\
  \times &\frac{1}{\sqrt{N_c}}\sum_{i=1}^{N_c} b_{s_q i}^\dagger(\xi P^+, \mathbf{k}+\xi\mathbf{P}) 
  d^\dagger_{s_{\bar{q}}i}((1-\xi)P^+, -\mathbf{k}+(1-\xi)\mathbf{P})| 0 \rangle,
\end{align}
where $P= (P^-, P^+, \mathbf{P})$ is the four momentum of the meson, $j$ and $m_j$ are respectively the total angular momentum and its magnetic projection. Here, $\psi_{s_q s_{\bar{q}}/h}^{(m_j)}(\mathbf{k}, \xi)$ are the valence-sector LFWFs, where $s_q$ and $s_{\bar{q}}$ represent the spins of the quark and the anti-quark respectively. For simplicity, we write $s=s_q$ and $\bar{s}=s_{\bar{q}}$. Explicitly, we expand the LFWF into the transverse and longitudinal basis functions with coefficients $\psi_h(n, m, l, s, \bar{s})$. In momentum space, 
\begin{align}\label{eq:LFWF_mom}
\psi_{s\bar{s}/h}(\mathbf{k},\xi)=\sum_{n,m,l} \psi_h(n,m,l,s,\bar{s}) \phi_{nm}(\frac{\mathbf{k}}{\sqrt{\xi(1-\xi)}})\chi_l(\xi),
\end{align}
where the basis functions are defined as 
\begin{align}
\begin{split}
  &\phi_{nm}(\mathbf{q}) = \frac{1}{\kappa}\sqrt{\frac{4\pi n!}{(n+|m|)!}} \Big ( \frac{|\mathbf{q}|}{\kappa} \Big)^{|m|} e^{-\frac{|\mathbf{q}|^2}{2 \kappa^2}} L_n^{|m|}(\frac{|\mathbf{q}|^2}{\kappa^2}) e^{i m \theta_q}\;, 
  \\
  &\chi_l(\xi; \alpha, \beta) =  \xi^{\frac{\beta}{2}}(1-\xi)^{\frac{\alpha}{2}}P_l^{(\alpha, \beta)}(2\xi-1) \sqrt{4\pi (2l+\alpha+\beta+1)}  \\
  &\qquad \qquad \qquad \times  \sqrt{\frac{\Gamma(l+1)\Gamma(l+\alpha+\beta+1)}{\Gamma(l+\alpha+1)\Gamma(l+\beta+1)}}\;. 
  \end{split}
\end{align}
Here, in the transverse direction, we use 2D harmonic oscillator basis functions, where $\theta_q = \mathrm{arg }(\mathbf{q})$, and $L_n^a(z)$ is the generalized Laguerre polynomial. Integers $n$ and $m$ are the principal quantum number for radial excitations and the orbital angular momentum projection quantum numbers, respectively. With this, the total angular momentum projection is $m_j = m+s+\bar{s}$. In the longitudinal basis direction, we use modified Jacobi polynomial $P_l^{(\alpha, \beta)}(z)$, where $l$ is the longitudinal quantum number. We have only two free parameters, $\kappa$ and $m_q$, from the model Hamiltonian when solving the meson spectroscopy; specifically, we use $\kappa=610$ MeV and $m_q=480$ MeV from fitting the $\rho$ meson mass and the pion mass~\cite{Qian:2020utg}. The confining strength $\kappa$ also serves as the harmonic oscillator scale parameter. The quantities $\alpha$ and $\beta$ are dimensionless basis variables, and in the limit of two equal quark masses ($m_q=m_{\bar{q}}$), $\alpha=\beta= 4m_q^2 /\kappa^2$.

By Fourier transformation, the LFWF in coordinate space becomes,
\begin{align}\label{eq:LFWF_pos}
\tilde{\psi}_{s\bar{s}/h}(\mathbf{r},\xi)=\sqrt{\xi(1-\xi)}\sum_{n,m,l} \psi_h(n,m,l,s,\bar{s}) \tilde{\phi}_{nm}(\sqrt{\xi(1-\xi)} \mathbf{r}) \chi_l(\xi),
\end{align}
where the 2D basis functions in position space are defined as
\begin{align}
      \tilde{\phi}_{nm}&(\mathbf{\rho}) = \kappa\sqrt{\frac{n!}{\pi(n+|m|)!}} \Big (\kappa |\mathbf{\rho}|  \Big)^{|m|} e^{-\frac{\kappa^2 |\mathbf{\rho}|^2}{2}} L_n^{|m|}(\kappa^2 |\mathbf{\rho}|^2) e^{i m \theta_q + i\pi(n+|m|/2)}. \label{eq:transverse_basis_pos}
\end{align}
Finally, the LFWFs are respectively normalized according to
\begin{subequations}\label{eq:normalization}
\begin{align}
  &\sum_{s, \bar{s}} \int_0^1 \frac{d\xi}{2\xi(1-\xi)} \int \frac{d^2\mathbf{k}}{(2\pi)^3}\psi_{s \bar{s}/h'}^{(m'_j)*}(\mathbf{k}, \xi) \psi_{s \bar{s}/h}^{(m_j)}(\mathbf{k}, \xi) = \delta_{hh'}\delta_{m_j m'_j}, \\
  &\sum_{s, \bar{s}} \int_0^1 \frac{d\xi}{4\pi} \int d^2\mathbf{r}\tilde{\psi}_{s \bar{s}/h'}^{(m'_j)*}(\mathbf{r}, \xi) \tilde{\psi}_{s \bar{s}/h}^{(m_j)}(\mathbf{r}, \xi) = \delta_{hh'}\delta_{m_j m'_j}.
\end{align}
\end{subequations}
With the basis functions defined, the LFWFs were numerically solved in a truncated basis~\cite{Qian:2020utg}, $N_\mathrm{max}=8$ and $L_\mathrm{max}=24$, giving sufficient energy resolution in both transverse and longitudinal directions: $2n+|m|+1 \leq N_\mathrm{max}$ and $\quad 0 \leq l \leq L_\mathrm{max}$. In the case of the pion, the leading contributions are summarized in Table~\ref{tab:basis_func_pion}. 
Although the $(n,m,l)=(0,0,0)$, $(0,1,0)$, and $(1,0,0)$ components contribute the most in the wave function, other higher order quantum numbers (400 in total) also play a significant role in shaping the complete wave function and in determining various observables. 

\begin{figure}
\centering
  \subfigure[\ $U_\mathrm{000}(r,\xi)$\label{fig:U000}]{\includegraphics[width=0.35\linewidth]{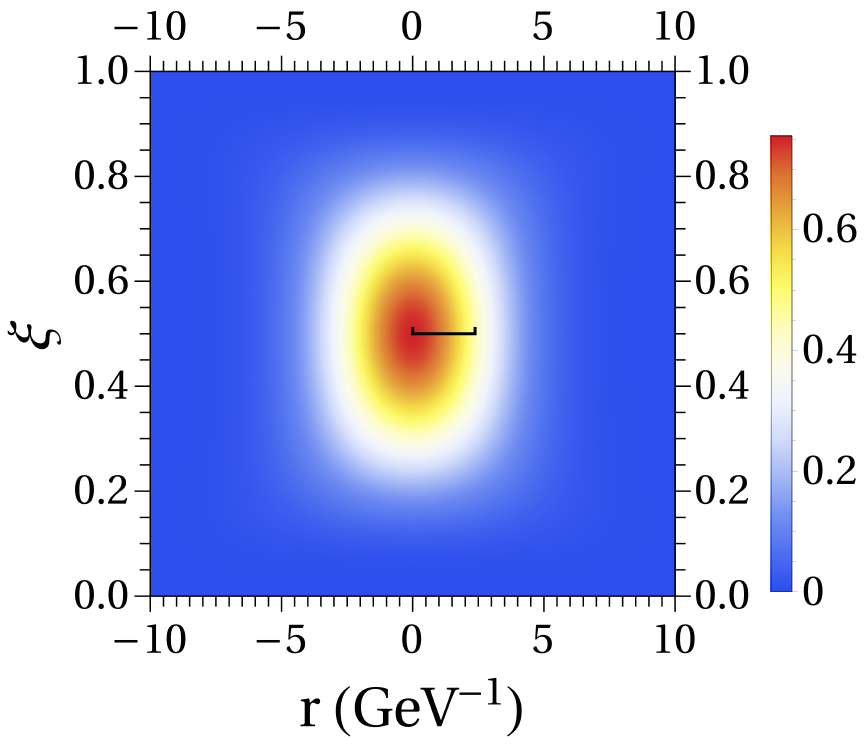}}
  \qquad
  \subfigure[\ $U_\mathrm{010}(r,\xi)$\label{fig:U010}]{\includegraphics[width=0.35\linewidth]{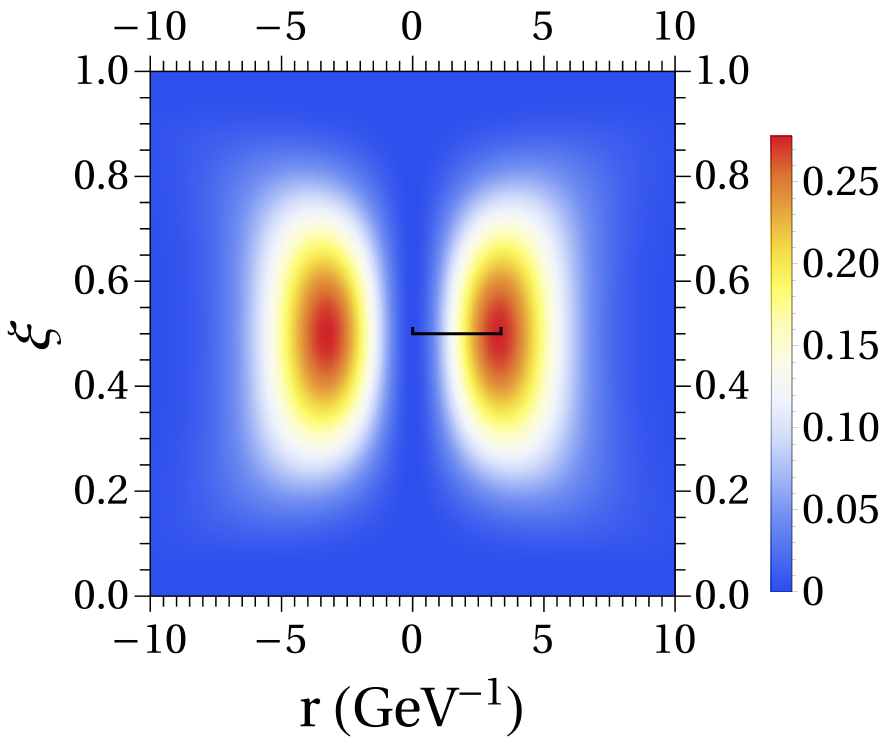}}
  \subfigure[\ $U_\mathrm{full}(r,\xi)$\label{fig:Ufull}]{\includegraphics[width=0.35\linewidth]{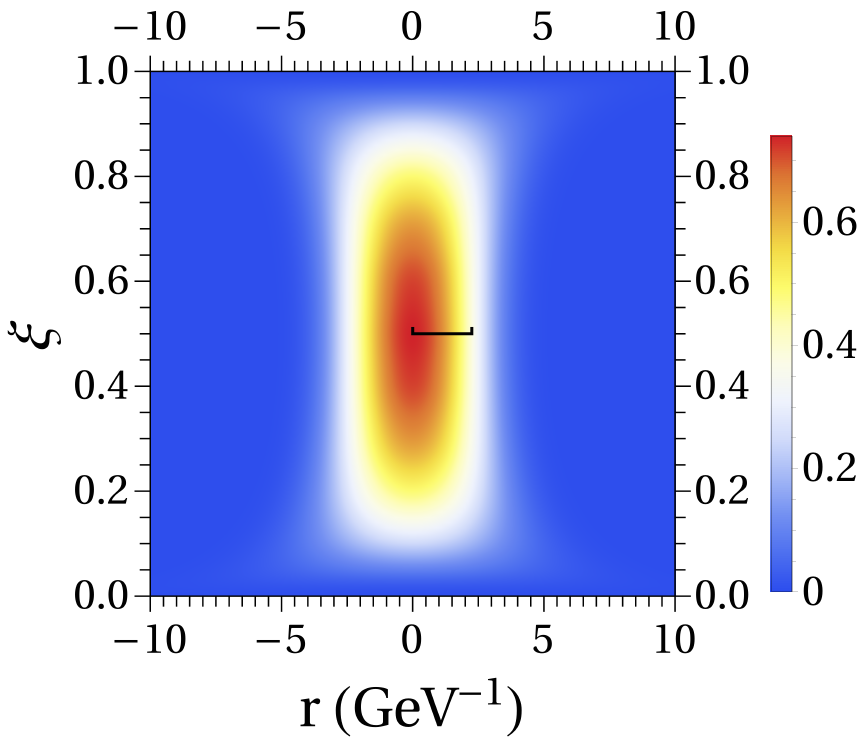}}
  \qquad
  \subfigure[\ $U_{\mathrm{top}\, 5\times 4}(r,\xi)$\label{fig:Utop5}]{\includegraphics[width=0.35\linewidth]{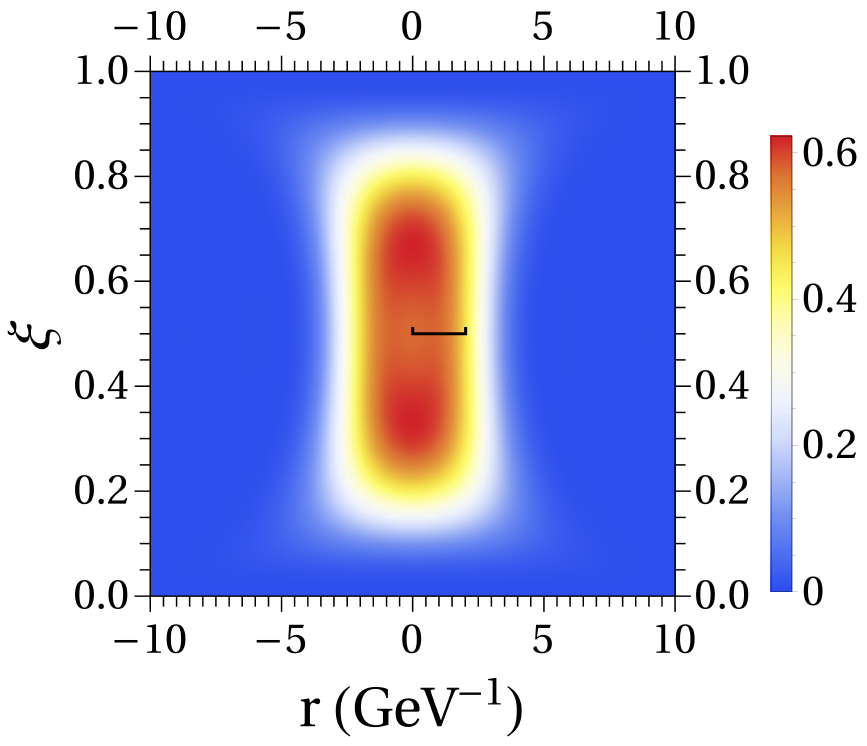}}
  \caption{Selected density plots of squared pion LFWFs according to eq.~\eqref{eq:sumspin_lfwf_squared}. The black segment shown on each plot represents the respective pion r.m.s. radius calculated using eq.~\eqref{eq:charge_radius_burkardt}.
  }
\label{fig:pion_LFWF_square_plot}
\end{figure}

Since spin sums of the squared LFWFs directly enter in the cross section calculations,  as in eq.~\eqref{eq:dsigmadbdo_pipi}, we define a convenient short-hand variable
\begin{align}\label{eq:sumspin_lfwf_squared}
    U(\mathbf{r}, \xi) &= \sum_{s\bar{s}} |\tilde{\psi}_{s\bar{s}}(\mathbf{r}, \xi)|^2= |\tilde{\psi}_{\uparrow\downarrow}(\mathbf{r}, \xi)|^2+|\tilde{\psi}_{\downarrow\uparrow}(\mathbf{r}, \xi)|^2+|\tilde{\psi}_{\downarrow\downarrow}(\mathbf{r}, \xi)|^2+|\tilde{\psi}_{\uparrow\uparrow}(\mathbf{r}, \xi)|^2,
\end{align} 
where $U(\mathbf{r}, \xi)$ = $U(|\mathbf{r}|, \xi)$ is azimuthally symmetric and normalized by $\int_0^1 \frac{d\xi}{4\pi} \int d^2\mathbf{r} U(\mathbf{r}, \xi) = 1$. In terms of the full and leading quantum numbers,
\begin{subequations}\label{eq:LFWF_Us}
\begin{align}
U_\mathrm{full}(r, \xi) &= \bigg|\sum_{nml}\tilde{\psi}_{nml\uparrow\downarrow}(r, \xi)\bigg|^2+\bigg|\sum_{nml}\tilde{\psi}_{nml\downarrow\uparrow}(r, \xi)\bigg|^2\\&+\bigg|\sum_{nml}\tilde{\psi}_{nml\downarrow\downarrow}(r, \xi)\bigg|^2+\bigg|\sum_{nml}\tilde{\psi}_{nml\uparrow\uparrow}(r, \xi)\bigg|^2,\nonumber\\
U_\mathrm{000}(r, \xi) &= \bigg|\frac{1}{\sqrt{2}}\tilde{\psi}_{000\uparrow\downarrow}(r, \xi)\bigg|^2+\bigg|\frac{-1}{\sqrt{2}}\tilde{\psi}_{000\downarrow\uparrow}(r, \xi)\bigg|^2 = |\tilde{\psi}_{000}(r, \xi)|^2,\\
U_\mathrm{010}(r, \xi) &= |\tilde{\psi}_{010\downarrow\downarrow}(r, \xi)|^2 =
U_\mathrm{0-10} = |\tilde{\psi}_{0-10\uparrow\uparrow}(r, \xi)|^2.
\end{align}
\end{subequations}
Selected density plots of $U(r, \xi)$ are included in Fig.~\ref{fig:pion_LFWF_square_plot}. Figures ~\ref{fig:U000}, ~\ref{fig:U010} and ~\ref{fig:Ufull} show $U_\mathrm{000}(r, \xi)$, $U_\mathrm{010}(r, \xi)$,  and the full LFWF squared, $U_\mathrm{full}(r, \xi)$, by summing contributions from all 400 basis functions. To study the sensitivity of physical observables on the number of basis states, we use the notation $U_{\mathrm{top}\, x\times4}$ to describe the leading spin-summed squared LFWFs with the first $x$ dominant components in each of the four spin configurations. For example, $U_{\mathrm{top}\, 100\times4} \equiv U_\mathrm{full}$ while $U_{\mathrm{top}\, 5\times4}$ has about 95\% of the LFWF, which means the probability of finding the pion in the top $5\times4$ basis states is about 95\%, assuming that the pion is 100\% in the full LFWF.
Comparing $U_\mathrm{full}$ and $U_{\mathrm{top}\, 5\times4}$, as shown in Fig.~\ref{fig:Ufull} and Fig.~\ref{fig:Utop5}, one can see the important contribution of higher-order basis states. In terms of physical observables, the root-mean-square charge radius (r.m.s. radius) is defined as the slope of the charge form factor $F_{ch}(Q^2)$ (FF) at zero momentum transfer,
\begin{align}
    \label{eq:charge_radius_ff}
    \braket{r_c^2} = -6\frac{\partial}{\partial Q^2} F_{ch}(Q^2)|_{Q\rightarrow 0},
\end{align}
which can also be equivalently and conveniently obtained by the Burkardt's impact parameter $\mathbf{b} = (1-x) \mathbf{r}$~\cite{Li:2017mlw} via
\begin{align}\label{eq:charge_radius_burkardt}
    \braket{r_c^2} =\frac{3}{2}\braket{\mathbf{b}^2}&=\frac{3}{2}\sum_{s,\bar{s}} \int_0^1 \frac{d\xi}{4\pi}\int d^2\mathbf{r} (1-\xi)^2 \mathbf{r}^2 \tilde{\psi}_{s \bar{s}/h'}^{*}(\mathbf{r}, \xi) \tilde{\psi}_{s \bar{s}/h}(\mathbf{r}, \xi)\notag\\
    &=\frac{3}{2}\int_0^1 \frac{d\xi}{4\pi}\int d^2\mathbf{r} (1-\xi)^2 \mathbf{r}^2 U(\mathbf{r}, \xi), 
\end{align}
where we summarize calculations of the r.m.s. radius of the pion in Table~\ref{tab:rms_pion}.

\begin{table*}
  \centering
  \caption{\label{tab:basis_func_pion}Leading basis functions for pions, ordered by LFWF coefficients $\psi_h(n,m,l,s,\bar{s})$. Recall that $m_j=m+s+\bar{s}$ by momentum conservation,and $s,\bar{s}$ are half integers. We can construct LFWFs using eq.~\eqref{eq:LFWF_pos} and eq.~\eqref{eq:LFWF_mom}.
  }
  \begin{ruledtabular}
    \begin{tabular}{c@{\hskip 0.01in}  c@{\hskip 0.1in}  c@{\hskip 0.1in}  c@{\hskip 0.1in}  c@{\hskip 0.1in} r@{\hskip 0.1in}}
    $n$ 
    & $m$
    & $l$ 
    & $s$
    & $\bar{s}$
    & $\psi_h(n,m,l,s,\bar{s})$
    \\[6pt] 
    \hline 
    0 & 0 & 0  & -1/2 & 1/2 & -0.442466  \\ 
    0 & 0 & 0  & 1/2 & -1/2 & 0.442466  \\  
    0 & 1 & 0  & -1/2 & -1/2 & -0.29865  \\   
    0 & -1 & 0  & 1/2 & 1/2 & -0.29865  \\   
    1 & 0 & 0  & -1/2 & 1/2 & 0.234734  \\   
    1 & 0 & 0  & 1/2 & -1/2 & -0.234734  \\   
    \end{tabular}
  \end{ruledtabular}
\end{table*}

\begin{table*}
  \centering
  \caption{\label{tab:rms_pion}The pion r.m.s. charge radius $\sqrt{\braket{r_c^2}}$ (in fm) calculated using eq.~\eqref{eq:charge_radius_burkardt} for various components of LFWFs, partial LFWFs and the full LFWFs. The charge radii are also plotted in Fig.~\ref{fig:pion_LFWF_square_plot} respectively. For comparison, the PDG data for the charge radius is 0.659 fm~\cite{Workman:2022ynf}. 
  }
  \begin{ruledtabular}
    \begin{tabular}{  l  c@{\hskip -0.05in}  c@{\hskip 0.01in}  c@{\hskip 0.01in} c@{\hskip 0.01in}  c@{\hskip 0.01in}  c@{\hskip 0.01in} 
    }
    & $(0,0,0)$ 
    & $(0,\pm 1, 0)$
    & $U_{\mathrm{top}\, 5\times4}$
    & $U_\mathrm{full}$
    \\[6pt]     \hline 
    $\sqrt{\braket{r_c^2}}$ 
    & 0.469 & 0.664 & 0.40 & 0.44 
    \\
    \end{tabular}
  \end{ruledtabular}
\end{table*}

\subsection{Transverse momentum anisotropies}
\label{sec:pionv2}

 By plugging the dipole cross section evaluated in sec.~\ref{eq:eval_sig_dipole} and the pion wave functions in the previous subsection into eq.~(\ref{eq:dsigmadbdo_pipi}), we now study transverse momentum anisotropies\footnote{The LO formula in eq.~(\ref{eq:sigppg}) produces non-vanishing even flow coefficients for $b>0$. We find that, e.g., $v_4$, with $v_4^x$ of opposite sign than $v_2^x$, is about one order of magnitude smaller than $v_2$ at the same $b$. In this section we only focus on the dominant flow coefficient $v_2$. } in high-energy pion-pion collisions. The results presented below are all obtained by using Monte Carlo methods 
 \cite{Hahn:2004fe,Hahn:2014fua} to carry out the integration over $\mathbf{r}_i$, $\xi_i$ and $\phi$ numerically.

 \begin{figure*}[htp!]
    \centering
    \includegraphics[width=0.6\textwidth]{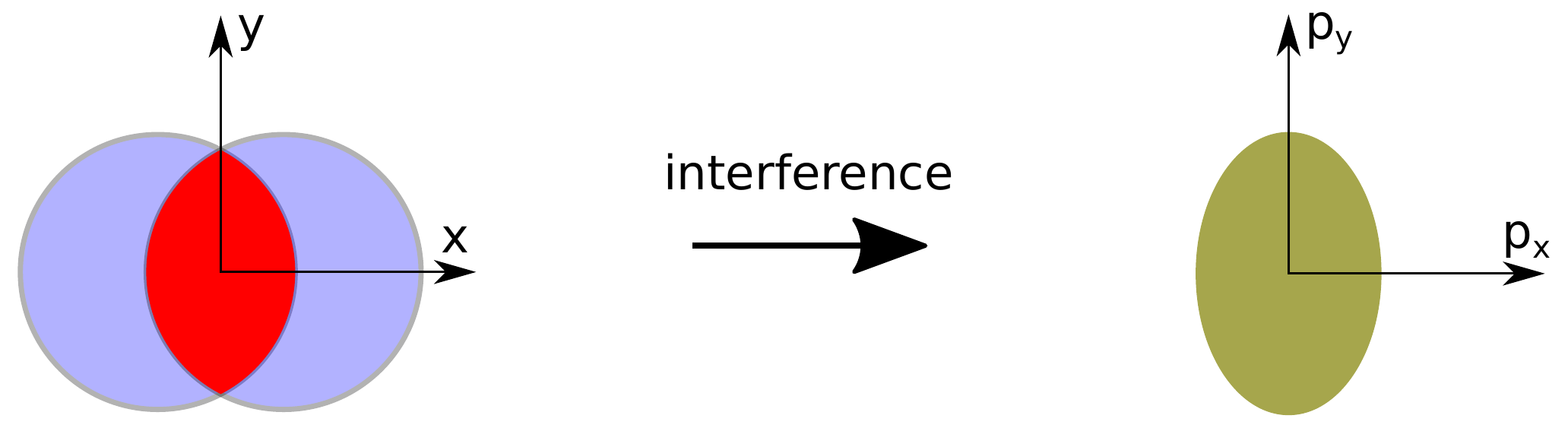}
    \caption{Collision geometry and transverse momentum anisotropies. According to our choice of the coordinates in eq.~(\ref{eq:choiceofb}), the reaction plane in the corresponding classical picture coincides with the $x$-$z$ plane. And the gluon is found to be more probably produced along the $y$-axis than the $x$-axis due to interference.
    }
    \label{fig:interference}
\end{figure*}

  For our choice of the coordinates in eq.~(\ref{eq:choiceofb}) in which the reaction plane coincides with the $x$-$z$ plane, one has $v_2^y=0$ due to reflection symmetry over the $x$-axis in the pion wave function squared (see eq.~\eqref{eq:vn_def} for the definition of $\mathbf{v}_n=(v_n^x, v_n^y)$). And we find that $v_2^x$ shown below is negative and, equivalently, the flow angle $\psi_2=\pi/2$, as illustrated in Fig.~\ref{fig:interference}. It is qualitatively different from the  expectation by naively extrapolating the classical hydrodynamic interpretation of elliptic flow in heavy-ion collisions~\cite{Ollitrault:1992bk} to hadron-hadron collisions. It would instead predict $\psi_2=0$, the same as the reaction plane angle. It is also different from the one-hit result in kinetic theory, which also has $\psi_2=0$~(see, e.g., ref.~\cite{Kurkela:2018ygx}).

\begin{figure*}[htp!]
    \centering
    \includegraphics[width=0.7\textwidth]{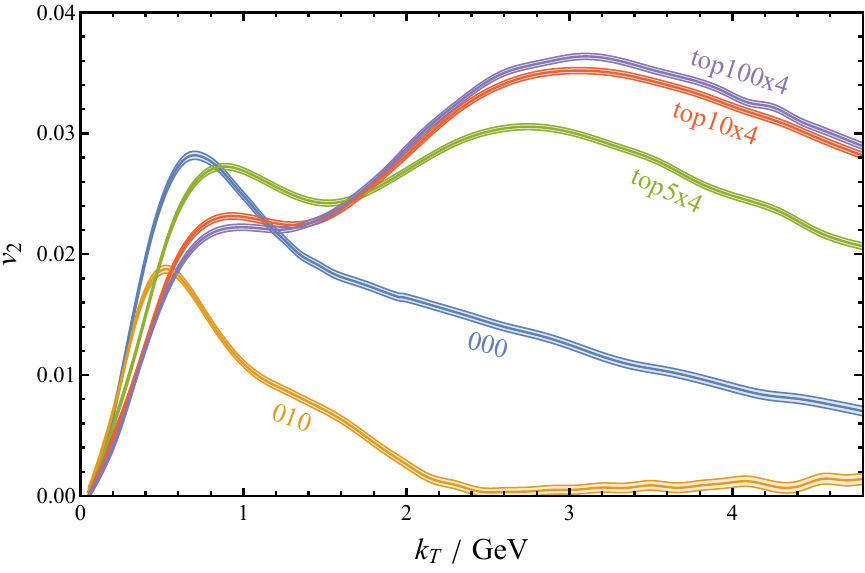}
    \caption{Sensitivity of $v_2$ on the shape of the pion light-front wave functions. The plot shows the results of $v_2$ for $U_{000}$, $U_{010}$, $U_{\mathrm{top}\, 5\times4}$, $U_{\mathrm{top}\, 10\times4}$ and $U_{\mathrm{top}\, 100\times4}$ (i.e., the full LFWF squared) at $b=0.2$~fm; see definitions of $U$s in eq.~\eqref{eq:LFWF_Us} and associated discussions. Here, the error bands are estimated as the uncertainties in Monte Carlo integration.
    }
    \label{fig:v2forU}
\end{figure*}

Fig.~\ref{fig:v2forU} shows the results of $v_2=|v_2^x|$  at $b=0.2$~fm by using one ($U_{000}$ and $U_{010}$), top 5 ($U_{\mathrm{top}\, 5\times4}$), top 10 ($U_{\mathrm{top}\, 10\times4}$) and top 100 ($U_{\text{full}}=U_{\mathrm{top}\, 100\times4}$) spin-summed, squared LFWFs. From this figure one can see that the shape of $v_2$ is quite sensitive to that of the wave functions, cf. Fig.~\ref{fig:pion_LFWF_square_plot}. Even $U_{\mathrm{top}\, 5\times4}$ produces a noticeably different shape of $v_2$ than the full one. Only if enough basis functions (top $10\times4$ or more) are included,  $v_2$ stabilizes (note the probability for the pion to occupy the basis states in $U_{\text{full}}$ not contained in $U_{\mathrm{top}\, 10\times4}$ is only about 1\%). For a comparison, the values of the r.m.s. charge radius for $U_{000}$, $U_{010}$, $U_{\mathrm{top}\, 5\times4}$ and $U_{\text{full}}$ are listed in Table~\ref{tab:rms_pion}, which do not show such a strong dependence on the shape of the wave functions (except $U_{010}$). Therefore, transverse momentum anisotropies could be used as a unique, stringent constraint on the shape of the hadron wave function, especially its high excited basis states.

\begin{figure*}[htp!]
    \centering
    \includegraphics[width=0.8\textwidth]{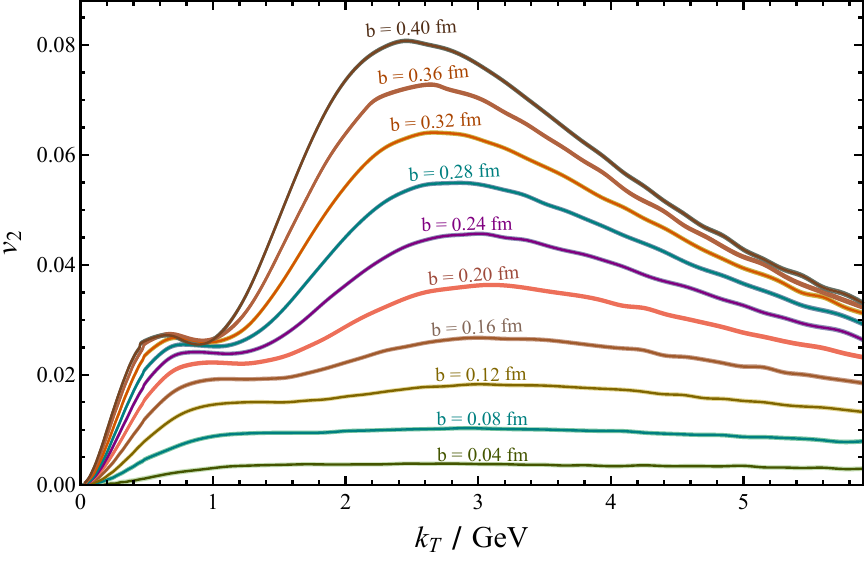}
    \caption{The elliptic flow $v_2$ as a function of $k_T$ at different impact parameters in pion-pion collisions. The collision geometry in the classical picture corresponding to each value of $b$ could be inferred from the pion r.m.s. charge radius $\sqrt{\braket{r_c^2}}$=0.44~fm. 
    }
    \label{fig:v2kT}
\end{figure*}

The pion $v_2$ at different impact parameters, calculated with the full LFWF, is shown as a function of $k_T$ in Fig.~\ref{fig:v2kT}. From central ($b=0.04$~fm) to peripheral ($b=0.4$~fm) collisions, $v_2$ always increases in the full range of $k_T\leq6$~GeV (except for the curve with $b=0.4$~fm at $0.6~\GeV \lesssim k_T \lesssim 1 \GeV$). At low $k_T$ the quadratic growth  survives in pion-pion collisions although the range of such a low-$k_T$ behavior is normally shortened compared to that in dipole-dipole collisions, as shown in Fig.~\ref{fig:v2b}. The qualitative behavior of $v_2$ for larger $k_T$ depends on the centrality of the collision. For $b>$0.1~fm, $v_2$ is characterized by a double-peak structure in which the global maximum locates in the range of $k_T=2-4$~GeV. 
Such a distinct structure becomes more pronounced at larger $b$. This qualitatively agrees with our observation in dipole-dipole scattering, cf. Fig.~\ref{fig:v2b}.

The double-peak structure in $v_2$ is a characteristic feature of the interference effect as discussed in sec.~\ref{sec:dipole_dipole_v2}. Let us take for example $b\sim \sqrt{\braket{r_c^2}}$, the pion r.m.s charge radius. In this case, $\sqrt{\braket{r_c^2}}=0.44$~fm is the only length scale in the problem. We find that the high $k_T$ behavior of $v_2$ in eq.~(\ref{eq:v2_large_kt}), although failing to predict the magnitude of $v_2$ for the shown range of $k_T$, gives a reasonable estimate of the locations of the two minima shown in Fig.~~\ref{fig:v2kT}: the first minimum locates around $k_T\sim 0.25\pi/\sqrt{\braket{r_c^2}}=0.36$~GeV while the second, around $k_T\sim 2.25\pi/\sqrt{\braket{r_c^2}}=3.2$~GeV. 
In addition, both peak locations mitigate to larger $k_T$ for decreasing $b$, as what one would expect from such an estimation, $k_{T,peak}\approx (0.25,2.25)\pi/b$.
That is, the peak locations are roughly determined by the hadron size (and the impact parameter).

\begin{figure*}[htp!]
    \centering
    \includegraphics[width=0.8\textwidth]{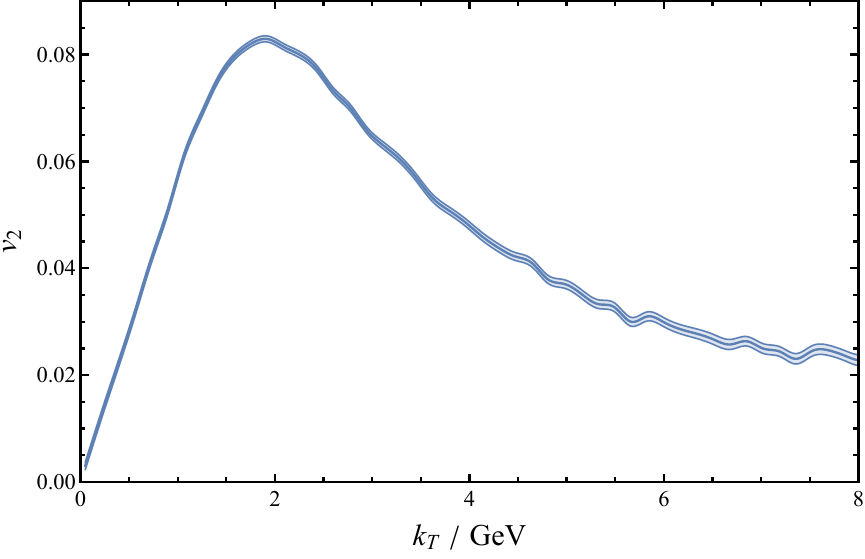}
    \caption{$v_2$ as a function of $k_T$ with $b$ integrated in pion-pion collisions.
    }
    \label{fig:v2kT_bIntegrated}
\end{figure*}

This paper is not aimed at a detailed discussion of flow phenomenology but to assess quantitatively the relative importance of the contributions to transverse momentum anisotropies from scattering of hadron valence quark skeletons in comparison with other initial-state and final-state effects~\cite{Nagle:2018nvi, Altinoluk:2020wpf}. As shown in Fig.~\ref{fig:v2kT}, with the only two parameters in the LFWFs fixed by light meson masses we find that the predicted $v_2$ at $b$ about the value of pion r.m.s. charge radius in soft gluon production is comparable to that observed in pp collisions~\cite{ATLAS:2015hzw, CMS:2016fnw, ATLAS:2017hap, ATLAS:2017rtr, CMS:2017kcs, ATLAS:2018ngv, ATLAS:2019wzn, CMS:2020qul} as well as theoretical results~\cite{dEnterria:2010xip, Bozek:2010pb, Habich:2015rtj, Weller:2017tsr, Zhao:2020pty, Dumitru:2010iy, Dusling:2012iga, Dusling:2013oia, Schenke:2014zha, Schenke:2016lrs, Iancu:2017fzn, Altinoluk:2020wpf}.  Note unlike heavy-ion collisions the impact parameter $b$ would not be well determined in pp collisions at the LHC~\cite{Wu:2021ril}, and we will need to find out its correlation with measurable quantities such as multiplicity in detailed phenomenological studies. In order to exemplify the effects of averaging over the impact parameter, we show the result of $v_2$ with $b$ integrated in Fig.~\ref{fig:v2kT_bIntegrated}. The maximum of the $b$-integrated $v_2$ ($\approx 0.08$), similar to that with $b\approx\sqrt{\braket{r_c^2}}$, is found to develop at lower $k_T\approx 2$~GeV while the double-peak structure is averaged out. The absence of the left peak is a combined effect of the large cross section with small $b$ and the large anisotropy with large $b$. Based on our observations in Figs.~\ref{fig:v2kT} and \ref{fig:v2kT_bIntegrated}, we believe that the interference effect from the emitters of valence (anti)quarks as discussed above would not be negligible in a comprehensive study of flow phenomenology. For example, if the events with large $b$ could be isolated, one could in principle observe the double-peak structure.

There are many issues to be addressed before we attempt to carry out phenomenological studies of collectivity in hadron-hadron collisions. First,  since the proton wave functions are already available~\cite{Liu:2022fvl,Hu:2022ctr,Xu:2022dbw,Xu:2022abw}, it would be intriguing and experimentally more relevant to generalize our calculations to proton-proton collisions. Second, in parton saturation/small-$x$ physics it was found that the contributions of the small-$x$ evolution could be significant in the description of collectivity~\cite{Levin:2011fb, Kovner:2011pe}. It, hence, would be of significance to study the effects of the small-$x$ evolution~\cite{Mueller:1993rr,Mueller:1994jq, Mueller:1994gb, Kovchegov:2005ur} on our LO results as well. Third, collectivity in multiple gluon production (higher-order cumulants) has been studied with~\cite{Agostini:2021xca} or without ~\cite{Blok:2017pui,Blok:2018xes} saturated dense gluons and transverse momentum anisotropies were found to persist. It would be important for us to carry out high-order calculations to study multi-particle correlations to confirm that observation. Fourth, either hadronization, the Local Parton-Hadron Duality~\cite{Azimov:1984np} or some physical observable such as the transverse energy needs to be introduced in order to compare with experimental data. Last but not least, the general formula for the impact-parameter dependent cross section in eq.~(\ref{eq:dsigmadbdosym}) is valid beyond the eikonal limit, which allows the exploration of non-eikonal effects. Such effects were shown to be sizable, e.g., at RHIC energies in CGC~\cite{Agostini:2019hkj, Agostini:2019avp, Agostini:2022ctk, Agostini:2022oge}. All these questions are left for future research.

\acknowledgments

We thank Nestor Armesto, Zhenyu Chen, Yuri Kovchegov, James P. Vary, Carlos Salgado, Yu Shi, Bo-Wen Xiao, and Xingbo Zhao for insightful  discussions. This work is supported by European Research Council project ERC-2018-ADG-835105 YoctoLHC; by Maria de Maetzu excellence program under project CEX2020-001035-M; by Spanish Research State Agency under project PID2020-119632GB- I00; and by Xunta de Galicia (Centro singular de investigación de Galicia accreditation 2019-2022), by European Union ERDF. H.Z. is supported by  the National Natural Science Foundation of China (NSFC) under Grant No. 12075136. 
B.W. acknowledges the support of the Ramón y Cajal program with the Grant No. RYC2021-032271-I.

\bibliographystyle{JHEP}
\bibliography{bulk.bib}
\end{document}